# COMPUTATIONAL METHODS TO SIMULTANEOUSLY COMPARE THE PREDICTIVE VALUES OF TWO DIAGNOSTIC TESTS WITH MISSING DATA: EM-SEM ALGORITHMS AND MULTIPLE IMPUTATION

J.A. Roldán-Nofuentes

Biostatistics, School of Medicine, University of Granada, 18016, Spain

Email: jaroldan@ugr.es

**Abstract.** Predictive values are measures of the clinical accuracy of a binary diagnostic test, and depend on the sensitivity and the specificity of the test and on the disease prevalence among the population being studied. This article studies hypothesis tests to simultaneously compare the predictive values of two binary diagnostic tests in the presence of missing data. The hypothesis tests were solved applying two computational methods: the EM and SEM algorithms and multiple imputation. Simulation experiments were carried out to study the sizes and the power of the hypothesis tests, giving some general rules of application. Two R programmes were written to apply each method, and they are available as supplementary material for the manuscript. The results were applied to the diagnosis of Alzheimer's disease.

# 1. Introduction

A diagnostic test is a medical test that is applied to a patient to determine the presence or absence of a certain disease, When the result of a diagnostic test is positive or negative, the diagnostic test is called a binary diagnostic test (*BDT*). The mammography for breast cancer is an example of a *BDT*. The clinical effectiveness of a BDT is measured in terms of two parameters: the positive predictive value (*PPV*) and the negative predictive value (*NPV*). The positive predictive value (*PPV*) is the probability of a patient having the disease when the result of the *BDT* is positive, and the negative predictive value (*NPV*) is the probability of the patient not having the disease when the result of the *BDT* is negative. The predictive values depend on the sensitivity (*Se*) and on the specificity (*Sp*) of the *BDT* and on the disease prevalence (*p*) among the population studied, i.e.

$$PPV = \frac{p \times Se}{p \times Se + q \times (1 - Sp)} \quad \text{and} \quad NPV = \frac{q \times Sp}{p \times (1 - Se) + q \times Sp} \qquad (1)$$

The predictive values are the parameters of the clinical accuracy of the *BDT*, since from the patient's point of view what is of interest it know the probability of having (or not having) the disease given a positive (negative) result for the diagnostic test. The predictive values of a *BDT* are estimated in relation to a gold standard (*GS*), which is a medical test which determines without any errors whether or not the patient has the disease. A biopsy for breast cancer is an example of a *GS*.

In clinical practice, the most common sample design to compare the predictive values of two *BDTs* is paired design [1, 2]. This type of design consists of applying the two *BDTs* to all of the individuals in a sample sized *n* whose disease status is known through the application of a *GS*. The comparison of the predictive values of two *BDTs* subject to paired design has been the subject of different studies in Statistics literature. Leisenring et al [3], Wang et al [4], Kosinski [5] and Tsou [6] have studied asymptotic methods to compare the two positive predictive values and the two negative predictive values independently, i.e. solving the two hypothesis tests $H_0: PPV_1 = PPV_2$ and $H_0: NPV_1 = NPV_2$ each one of them to an $\alpha$ error. The Kosinski method has a better asymptotic performance (in terms of type I error and power) than the methods of Leisenring et al and of Wang et al. The method of Tsou leads to the same results as the Kosinski method. Roldán Nofuentes et al [7] studied a global hypothesis test to simultaneously compare the predictive values of two *BDTs*, i.e. solving the global hypothesis test $H_0: (PPV_1 = PPV_2 \text{ and } NPV_1 = NPV_2)$ vs $H_1: (PPV_1 \neq PPV_2 \text{ and/or } NPV_1 \neq NPV_2)$, and proposed a method based on chi-squared distribution and multiple comparisons. These authors have demonstrated that the comparison of the positive (negative) predictive values of two *BDTs* subject to a paired design must be carried out simultaneously, solving the global hypothesis test $H_0: (PPV_1 = PPV_2 \text{ and } NPV_1 = NPV_2)$. They have also demonstrated that the comparison of the predictive values is made independently, i.e. solving the tests $H_0: PPV_1 = PPV_2$ and $H_0: NPV_1 = NPV_2$ each one of them to an $\alpha$ error, the results may be mistaken. In Appendix *A* the aforementioned methods are summarized.

When assessing or comparing parameters de *BDTs* it is not common for the *GS* to be applied to all of the individuals in the sample, leading to the problem known as partial disease verification [8, 9]. Therefore, if the *GS* consists of a costly test or one which means some risk for the individual, then it will not be applied to all of the individuals in the sample, and consequently the true disease status is unknown for a subgroup of individuals. When comparing parameters of two *BDTs* in the presence of partial disease verification, it is common to assume that the verification process is missing at random (*MAR*). This assumes that the process to verify the disease status of an individual through the application of the *GS* only depends on the results of the two *BDTs* and it does not depend on the disease status of the individual. Subject to the *MAR* assumption, there are many different studies in Statistics literature which compare parameters of two or more *BDTs* in the presence of partial verification. Zhou [9] studied a hypothesis test to compare the sensitivities (specificities) of two *BDTs* applying the method of maximum likelihood. Roldán Nofuentes and Luna [10] studied the individual comparison of the predictive values of two *BDTs* applying maximum likelihood. Marín-Jiménez and Roldán Nofuentes [11] extended the study of Roldán-Nofuentes and Luna [10] to the case of more than two *BDTs* comparing the predictive values simultaneously. Harel and Zhou [12] compared the sensitivities (specificities) of two *BDTs* through confidence intervals applying multiple imputation, and recommending the use of multiple imputation along with the continuity-corrected Wald interval to compare the two sensitivities (specificities). Roldán-Nofuentes and Luna [13] compared the sensitivities and the specificities independently, as well as the predictive values, of two *BDTs* applying the *EM* and *SEM* algorithms.

In this article, hypothesis tests are studied to simultaneously compare the predictive values of two *BDTs* when in the presence of partial disease verification the missing data mechanism is *MAR*, applying two computational methods: the *EM* and *SEM* algorithms and multiple imputation. Therefore, with both types of computational methods we seek to solve the global hypothesis global test to simultaneously compare the positive predictive values and the negatives predictive values of the two *BDTs*, i.e.

$$H_0 : PPV_1 = PPV_2 \text{ and } NPV_1 = NPV_2$$
$$H_1 : PPV_1 \neq PPV_2 \text{ and/or } NPV_1 \neq NPV_2. \qquad (2)$$

In Section 2, we solve the global test applying the *EM* and *SEM*, and in Section 3 the same test is solved applying multiple imputation. In Section 4, simulation experiments are carried out to study the Type I errors and the power of the previous tests with each one of the two methods (*EM-SEM* algorithms and multiple imputation), and some general rules of application are given. In Section 5, two programmes written in R which allow us to solve the problem posed applying each one of the two computational methods. In Section 6, the results were applied to a real example of the diagnosis of Alzheimer's disease, and in Section 7 the results obtained are discussed.

## 2. *EM* and *SEM* algorithms

Let us consider two *BDTs* that are applied to all of the individuals in a random sample sized $n$, and let us also consider a *GS* that is only applied to a subgroup of the sample. This situation leads to the observed frequencies in Table 1 (observed frequencies in the presence of partial verification), where the $T_h$ models the result of the $h$th *BDT* ($T_h = 1$ when the result is positive and $T_h = 0$ when it is negative), the variable $V$ models the verification process ($V = 1$ when the disease status of an individual is verified with the *GS* and $V = 0$ when it is not), and the variable $D$ models the result of the *GS* ($D = 1$ when the individual verified has the disease and $D = 0$ when this individual does not). In Table 1 (observed frequencies in the presence of partial verification), $a_{ij}$ is the number of individuals with the disease among whom $T_1 = i$ and $T_2 = j$, $b_{ij}$ is the number of individuals without the disease among whom $T_1 = i$ and $T_2 = j$, and $c_{ij}$ is the number de individuals with an unknown disease status among whom $T_1 = i$ and $T_2 = j$, with $i, j = 0, 1$. Sean $a = \sum_{i,j=0}^{1} a_{ij}$, $b = \sum_{i,j=0}^{1} b_{ij}$, $c = \sum_{i,j=0}^{1} c_{ij}$, $n_{ij} = a_{ij} + b_{ij} + c_{ij}$ and $n = \sum_{i,j=0}^{1} n_{ij}$.

For the $h$th *BDT*, let $Se_h = P(T_h = 1 | D = 1)$, $Sp_h = P(T_h = 0 | D = 0)$, $PPV_h = P(D = 1 | T_h = 1)$ and $NPV_h = P(D = 0 | T_h = 0)$ be the sensitivity, the specificity, the positive predictive value and the negative predictive value respectively, with $h = 1, 2$. Let $p = P(D = 1)$ be the disease prevalence and $q = 1 - p = P(D = 0)$. From the expressions (1) each *Se* and *Sp* is written, in terms of the predictive values and of $p$, as

$$Se_h = \frac{PPV_h(NPV_h - q)}{pY_h} \quad \text{and} \quad Sp_h = \frac{NPV_h(PPV_h - p)}{qY_h}, \tag{3}$$

where $Y_h = PPV_h + NPV_h - 1$.

In the presence of partial disease verification, the verification probabilities are designed as $\lambda_{ijk} = P(V = 1 | T_1 = i, T_2 = j, D = k)$, i.e. $\lambda_{ijk}$ is the probability of verifying the *GS* the disease status of an individual for whom $T_1 = i$, $T_2 = j$ and $D = k$, with $i, j, k = 0, 1$. Assuming that the missing data mechanism is *MAR*, i.e. that the probability of verifying the disease status of an individual only depends on the results of the two *BDTs* and does not depend on the result of the *GS*, it is verified that

$$\lambda_{ijk} = \lambda_{ij} = P(V = 1 | T_1 = i, T_2 = j). \tag{4}$$

Supposing the *MAR* assumption, the data from Table 1 (observed frequencies in the presence of partial verification) are the product of a multinomial distribution sized $n$ and the probabilities are written in terms of the predictive values as

$$\xi_{ij} = P(V = 1, D = 1, T_1 = i, T_2 = j) =$$
$$p\lambda_{ij}\left[\frac{PPV_1^i(v_1 - q)^i(PPV_1 + p)^{1-i}(1 - NPV_1)^{1-i}}{p^i p^{1-i} Y_1^i Y_1^{1-i}} \times \frac{PPV_2^j(NPV_2 - q)^j(PPV_2 + p)^{1-j}(1 - NPV_2)^{1-j}}{p^j p^{1-j} Y_2^j Y_1^{1-j}} + \delta_{ij}\varepsilon_1\right],$$

$$\psi_{ij} = P(V = 1, D = 0, T_1 = i, T_2 = j) = \tag{5}$$
$$q\lambda_{ij}\left[\frac{(1 - PPV_1)^i(NPV_1 - q)^i(PPV_1 - p)^{1-i} NPV_1^{1-i}}{q^i q^{1-i} Y_1^i Y_1^{1-i}} \times \frac{(1 - PPV_2)^j(NPV_2 - q)^j(PPV_2 - p)^{1-j} NPV_2^{1-j}}{q^j q^{1-j} Y_2^j Y_2^{1-j}} + \delta_{ij}\varepsilon_0\right],$$

$$\zeta_{ij} = P(V = 0, T_1 = i, T_2 = j) = \frac{1 - \lambda_{ij}}{\lambda_{ij}}(\xi_{ij} + \psi_{ij}),$$

where

$$\varepsilon_1 = \frac{PPV_1 PPV_2 (NPV_1 - q)(NPV_2 - q)(\alpha_1 - 1)}{p^2 Y_1 Y_2}$$

and

$$\varepsilon_0 = \frac{(1 - PPV_1)(1 - PPV_2)(NPV_1 - q)(NPV_2 - q)(\alpha_0 - 1)}{q^2 Y_1 Y_2}.$$

with $\delta_{ij} = 1$ if $i = j$ and $\delta_{ij} = -1$ if $i \neq j$. Parameters $\alpha_1$ and $\alpha_0$ are the covariances [14] between the two *BDTs* when $D = 1$ and when $D = 0$ respectively, verifying that

$$1 \leq \alpha_1 \leq \frac{1}{\max\left\{\dfrac{PPV_1(NPV_1 - q)}{pY_1}, \dfrac{PPV_2(NPV_2 - q)}{pY_2}\right\}}$$

and

$$1 \leq \alpha_0 \leq \frac{1}{\max\left\{\left(1 - \dfrac{NPV_1(PPV_1 - p)}{qY_1}\right), \left(1 - \dfrac{NPV_2(PPV_2 - p)}{qY_2}\right)\right\}}.$$

If $\alpha_1 = \alpha_0 = 1$, then the two *BDTs* are conditionally independent on the disease, a situation which is not realistic in practice, and therefore it must be verified that $\alpha_1 > 1$ and/or $\alpha_0 > 1$.

An *EM* algorithm is then proposed to estimate the predictive values of the two *BDTs*.

*2.1. EM algorithm*

In the $3 \times 4$ table of observed frequencies, the missing information is the true disease status of the individuals who are not verified with the *GS*, i.e. the missing data is the value of the variable $D$ for the individuals among whom $V = 0$. This information is reconstructed in the *E* step of the algorithm and in the *M* step the values of the maximum likelihood estimators are imputed. Let us assume that that among the $c_{ij}$ individuals who are not verified $(V = 0)$, $d_{ij}$ have the disease and $c_{ij} - d_{ij}$ do not have it, with $i, j = 0, 1$. Then the table of observed frequencies can be expressed in the form of a $2 \times 4$ table with frequencies $a_{ij} + d_{ij}$ for $D = 1$ and $b_{ij} + c_{ij} - d_{ij}$ for $D = 0$ (Table 1: Complete data). Let $\boldsymbol{\theta} = (PPV_1, NPV_1, PPV_2, NPV_2, p, \alpha_1, \alpha_0)$ be the vector of parameters. From the table of complete data, the log-likelihood function based on *n* individuals is

$$l(\boldsymbol{\theta}) = \sum_{i,j=0}^{1}(a_{ij}+d_{ij})\log(\phi_{ij}) + \sum_{i,j=0}^{1}(b_{ij}+c_{ij}-d_{ij})\log(\varphi_{ij}), \qquad (6)$$

where $\phi_{ij} = P(T_1=i, T_2=i, D=1)$ and $\varphi_{ij} = P(T_1=i, T_2=i, D=0)$. These probabilities are written in terms of the predictive values as

$$\phi_{11} = \frac{\alpha_1 PPV_1 PPV_2 (NPV_1 - q)(NPV_2 - q)}{pY_1 Y_2},$$

$$\phi_{10} = \frac{PPV_1 (NPV_1 - q)[pY_2 - \alpha_1 PPV_2 (NPV_2 - q)]}{pY_1 Y_2},$$

$$\phi_{01} = \frac{PPV_2 (NPV_2 - q)[pY_1 - \alpha_1 PPV_1 (NPV_1 - q)]}{pY_1 Y_2},$$

$$\phi_{00} = \frac{PPV_1 (NPV_1 - q)\{p[1 - NPV_2 + PPV_2(\alpha_1 - 1)] - \alpha_1 PPV_2 (1 - NPV_2)\}}{pY_1 Y_2} - \frac{p(p - PPV_2)(1 - NPV_2)Y_1}{pY_1 Y_2},$$

$$\varphi_{11} = \frac{\alpha_0 (1 - PPV_1)(1 - PPV_2)(NPV_1 - q)(NPV_2 - q)}{qY_1 Y_2},$$

$$\varphi_{10} = \frac{(1 - PPV_1)(NPV_1 - q)[qY_2 - \alpha_0 (1 - PPV_2)(NPV_2 - q)]}{qY_1 Y_2},$$

$$\varphi_{01} = \frac{(1 - PPV_2)(NPV_2 - q)[qY_1 - \alpha_0 (1 - PPV_1)(NPV_1 - q)]}{qY_1 Y_2},$$

$$\varphi_{00} = \frac{\alpha_0 (1 - PPV_1)(1 - PPV_2)(NPV_1 - q)(NPV_2 - q) - q^2 Y_1 Y_2}{qY_1 Y_2} + \qquad (7)$$

$$\frac{(1 - PPV_2)(PPV_1 - p)NPV_1 + [p(2NPV_1 - 1) - PPV_1(NPV_1 - p) - PPV_2 Y_1]NPV_2}{qY_1 Y_2}$$

The components of the vector $\boldsymbol{\theta}$ are going to be estimated applying the *EM* algorithm. Therefore, let us suppose that $d_{ij}^{(k)}$ is the value of $d_{ij}$ in the *k*th iteration of the *EM* algorithm, and $d^{(k)} = \sum_{i,j=0}^{1} d_{ij}^{(k)}$. The values of the *MLEs* in the *k*th iteration are calculated through the following equations

$$\widehat{PPV}_1^{(k)} = \frac{\sum_{j=0}^{1}\left(a_{1j}+d_{1j}^{(k)}\right)}{\sum_{j=0}^{1}n_{1j}}, \quad \widehat{PPV}_2^{(k)} = \frac{\sum_{i=0}^{1}\left(a_{i1}+d_{i1}^{(k)}\right)}{\sum_{i=0}^{1}n_{i1}},$$

$$\widehat{NPV}_1^{(k)} = \frac{\sum_{j=0}^{1}\left(b_{0j}+c_{0j}-d_{0j}^{(k)}\right)}{\sum_{j=0}^{1}n_{0j}}, \quad \widehat{NPV}_2^{(k)} = \frac{\sum_{i=0}^{1}\left(b_{i0}+c_{i0}-d_{i0}^{(k)}\right)}{\sum_{i=0}^{1}n_{i0}}, \quad (8)$$

$$\hat{p}^{(k)} = \frac{a+d^{(k)}}{n}, \quad \hat{\alpha}_1^{(k)} = \frac{\left(a+d^{(k)}\right)\left(a_{11}+d_{11}^{(k)}\right)}{\left[\sum_{i=0}^{1}\left(a_{i1}+d_{i1}^{(k)}\right)\right]\left[\sum_{j=0}^{1}\left(a_{1j}+d_{1j}^{(k)}\right)\right]},$$

$$\hat{\alpha}_0^{(k)} = \frac{\left(b+c-d^{(k)}\right)\left(b_{11}+c_{11}-d_{11}^{(k)}\right)}{\left[\sum_{i=0}^{1}\left(b_{i1}+c_{i1}-d_{i1}^{(k)}\right)\right]\left[\left\{\sum_{j=0}^{1}\left(b_{1j}+c_{1j}-d_{1j}^{(k)}\right)\right\}\right]}.$$

The estimators in the $(k+1)$th iteration of the algorithm are calculated applying equations (8) substituting the superindex $k$ with $k+1$, where

$$d_{ij}^{(k+1)} = c_{ij}\frac{\hat{\phi}_{ij}^{(k)}}{\hat{\phi}_{ij}^{(k)}+\hat{\varphi}_{ij}^{(k)}}, \quad i,j=0,1,$$

when $\hat{\phi}_{ij}^{(k)}$ and $\hat{\varphi}_{ij}^{(k)}$ are the estimators of the probabilities $\phi_{ij}$ and $\varphi_{ij}$ in the $k$th iteration of the algorithm, and which are calculated substituting in the expressions of $\phi_{ij}$ and $\varphi_{ij}$ (equations (7)) the parameters with their respective estimators obtained in the $k$th iteration. As the initial value $d_{ij}^{(0)}$ one can take any value between 0 and $u_{ij}$. The *EM* algorithm stops when the difference between the values of the log-likelihood functions of two consecutive iterations is lower than a value $\delta$, for example $\delta=10^{-10}$ or $\delta=10^{-12}$. If the *EM* algorithm has converged in $K$ iterations, we denote through $\hat{\boldsymbol{\theta}} = \left(\widehat{PPV}_1, \widehat{NPV}_1, \widehat{PPV}_2, \widehat{NPV}_2, \hat{p}, \hat{\alpha}_1, \hat{\alpha}_0\right)$ the final estimators obtained.

The variances-covariances of $\hat{\boldsymbol{\theta}}$ are then estimated applying the *SEM* algorithm [15].

*2.2. SEM Algorithm*

The estimation of the matrix of asymptotic variances-covariances $\hat{\boldsymbol{\theta}}$ can be obtained applying the *SEM* algorithm [15], which is a computational method which estimates the variances-covariances matrix of a vector of estimators from the calculations performed in the application of the *EM* algorithm. Let $\Sigma_{\hat{\boldsymbol{\theta}}}$ be the variances-covariances matrix of $\hat{\boldsymbol{\theta}} = \left(\widehat{PPV_1}, \widehat{NPV_1}, \widehat{PPV_2}, \widehat{NPV_2}, \hat{p}, \hat{\alpha}_1, \hat{\alpha}_0\right)$, Dempster et al [16] demonstrated that

$$\Sigma_{\hat{\boldsymbol{\theta}}} = I_{oc}^{-1}\left(I - DM\right)^{-1}, \qquad (9)$$

where $I$ is the identity matrix and $DM = I_{mis}I_{oc}^{-1}$, when $I_{oc}$ is the Fisher information matrix of the complete data and $I_{mis}$ is the Fisher information matrix of the missing data. The *SEM* algorithm consists of three phases: 1) assessment of the matrix $I_{oc}^{-1}$, 2) assessment of the matrix *DM*, and 3) assessment of the matrix $\Sigma_{\hat{\boldsymbol{\theta}}}$. The main phase is to calculate the elements of the *DM* matrix. The following three phases are then analysed.

The first phase consists of assessing the $I_{oc}^{-1}$. This matrix is the inverse of the Fisher information matrix of the complete data, i.e.

$$I_{oc} = -\frac{\partial^2 l(\boldsymbol{\theta})}{\partial \theta_i \partial \theta_j},$$

where $l(\boldsymbol{\theta})$ is the function (6) and each $\theta_i$ is one of the parameters $\boldsymbol{\theta} = \left(PPV_1, NPV_1, PPV_2, NPV_2, p, \alpha_1, \alpha_0\right)$. This matrix is calculated from the last table after the application of the *EM* algorithm, substituting the parameters with their corresponding estimations obtained in the last iteration of the *EM* algorithm. If the *EM* algorithm has converged in $K$ iterations, then the frequencies of the last $2 \times 4$ table are $a_{ij} + d_{ij}^{(K)}$ for $D = 1$ and $b_{ij} + c_{ij} - d_{ij}^{(K)}$ for $D = 0$.

The second part of the *SEM* algorithm consists of calculating the *DM* matrix. The elements of this matrix, denoted as $r_{ij}$, $i,j = 1,...,7$, are obtained applying the following algorithm:

INPUT: $\hat{\boldsymbol{\theta}}$ and $\boldsymbol{\theta}^{(k)} = \left( PPV_1^{(k)}, NPV_1^{(k)}, PPV_2^{(k)}, NPV_2^{(k)}, p^{(k)}, \alpha_1^{(k)}, \alpha_0^{(k)} \right)$.

Step 1: Calculate $\boldsymbol{\theta}^{(k+1)} = \left( PPV_1^{(k+1)}, NPV_1^{(k+1)}, PPV_2^{(k+1)}, NPV_2^{(k+1)}, p^{(k+1)}, \alpha_1^{(k+1)}, \alpha_0^{(k+1)} \right)$ applying the *EM* algorithm proposed in Section 2.1.

Step 2: Obtain the vectors

$$\boldsymbol{\theta}_1^{(k)} = \left( PPV_1^{(k)}, \widehat{NPV_1}, \widehat{PPV_2}, \widehat{NPV_2}, \hat{p}, \hat{\alpha}_1, \hat{\alpha}_0 \right),$$
$$\boldsymbol{\theta}_2^{(k)} = \left( \widehat{PPV_1}, NPV_1^{(k)}, \widehat{PPV_2}, \widehat{NPV_2}, \hat{p}, \hat{\alpha}_1, \hat{\alpha}_0 \right),$$
$$\boldsymbol{\theta}_3^{(k)} = \left( \widehat{PPV_1}, \widehat{NPV_1}, PPV_2^{(k)}, \widehat{NPV_2}, \hat{p}, \hat{\alpha}_1, \hat{\alpha}_0 \right),$$
$$\boldsymbol{\theta}_4^{(k)} = \left( \widehat{PPV_1}, \widehat{NPV_1}, \widehat{PPV_2}, NPV_2^{(k)}, \hat{p}, \hat{\alpha}_1, \hat{\alpha}_0 \right),$$
$$\boldsymbol{\theta}_5^{(k)} = \left( \widehat{PPV_1}, \widehat{NPV_1}, \widehat{PPV_2}, \widehat{NPV_2}, p^{(k)}, \hat{\alpha}_1, \hat{\alpha}_0 \right),$$
$$\boldsymbol{\theta}_6^{(k)} = \left( \widehat{PPV_1}, \widehat{NPV_1}, \widehat{PPV_2}, \widehat{NPV_2}, \hat{p}, \alpha_1^{(k)}, \hat{\alpha}_0 \right),$$
$$\boldsymbol{\theta}_7^{(k)} = \left( \widehat{PPV_1}, \widehat{NPV_1}, \widehat{PPV_2}, \widehat{NPV_2}, \hat{p}, \hat{\alpha}_1, \alpha_0^{(k)} \right),$$

and for each one of these seven vectors run the first iteration of the *EM* algorithm taking $\boldsymbol{\theta}_i^{(k)}$ as the initial value of $\boldsymbol{\theta}$, and obtain the vectors $\tilde{\boldsymbol{\theta}}_1^{(k+1)}$, $\tilde{\boldsymbol{\theta}}_2^{(k+1)}$,..., $\tilde{\boldsymbol{\theta}}_7^{(k+1)}$.

Step 3: Calculating the elements of the *DM* matrix as

$$r_{ij}^{(k)} = \frac{\tilde{\theta}_{ij}^{(k+1)} - \hat{\theta}_j}{\theta_i^{(k)} - \hat{\theta}_i}, \quad i,j = 1,...,7,$$

where $\tilde{\theta}_{ij}^{(k+1)}$ is the *j*th component of vector $\tilde{\boldsymbol{\theta}}_i^{(t+1)}$, $\theta_i^{(k)}$ is the *i*th component of vector $\boldsymbol{\theta}^{(k)}$ and $\hat{\theta}_i$ is the *i*th component of vector $\hat{\boldsymbol{\theta}}$.

OUTPUT: $\boldsymbol{\theta}^{(k+1)}$ and $r_{ij}^{(k)}$, $i,j = 1,...,7$.

This algorithm is repeated until $\left| r_{ij}^{(k+1)} - r_{ij}^{(k)} \right| \leq \sqrt{\delta}$ [15]. Consequently, the lower $\delta$ is, the lower the numerical errors that are made when calculating the elements of the *DM* matrix, thus making fewer numerical errors in the estimation of the variances-covariances matrix $\Sigma_{\hat{\theta}}$.

The last phase of the *SEM* algorithm consists of estimating the variances-covariances matrix $\Sigma_{\hat{\theta}}$ applying equation (9). The estimated variances-covariances matrix is not normally symmetrical due to the numerical errors made in the calculation de la matrix *DM*. The assessment of $\hat{\Sigma}_{\hat{\theta}}$ is performed calculating the matrix $\Delta\hat{\Sigma}_{\hat{\theta}} = \hat{I}_{oc}^{-1} DM (I - DM)^{-1}$, a matrix which represents the increase in the variances-covariances estimated owing to the missing information. The smaller the value of $\delta$ is, the more symmetrical the matrix $\Delta\hat{\Sigma}_{\hat{\theta}}$ will be, and therefore the more symmetrical $\hat{\Sigma}_{\hat{\theta}}$ will be. Thus, the problem of the asymmetry of $\hat{\Sigma}_{\hat{\theta}}$ is solved taking a value a very small value of $\delta$ [15].

*2.3. Global Test*

The global hypothesis test (2) to simultaneously compare the predictive values of the two *BDTs* is equivalent to

$$H_0 : \gamma\eta = 0 \text{ vs } H_1 : \gamma\eta \neq 0. \qquad (10)$$

where $\eta = (PPV_1, NPV_1, PPV_2, NPV_2)^T$ the vector whose components are the predictive values of the two *BDTs* and $\gamma$ is a complete range matrix sized $2 \times 4$ whose elements are known constants, i.e.

$$\gamma = \begin{pmatrix} 1 & 0 & -1 & 0 \\ 0 & 1 & 0 & -1 \end{pmatrix}.$$

Applying the multivariate central limit theorem it is verified that $\sqrt{n}(\hat{\eta} - \eta) \xrightarrow[n \to \infty]{} N(0, \Sigma_\eta)$ where $\Sigma_\eta$ is the variance-covariance matrix of $\eta$. Then, the statistic

$$Q^2 = \hat{\boldsymbol{\eta}}^T \boldsymbol{\gamma}^T \left( \boldsymbol{\gamma} \hat{\boldsymbol{\Sigma}}_{\hat{\boldsymbol{\eta}}} \boldsymbol{\gamma}^T \right)^{-1} \boldsymbol{\gamma} \hat{\boldsymbol{\eta}}$$

is distributed according to a Hotelling's $T$-squared distribution dimension 2 and $n$ degrees of freedom, where 2 is the dimension of the vector $\hat{\boldsymbol{\eta}}$. When $n$ is large, the statistic $Q^2$ is distributed according to a central chi-squared distribution with 2 degrees of freedom when the null hypothesis is true, i.e.

$$Q^2 = \hat{\boldsymbol{\eta}}^T \boldsymbol{\gamma}^T \left( \boldsymbol{\gamma} \hat{\boldsymbol{\Sigma}}_{\hat{\boldsymbol{\eta}}} \boldsymbol{\gamma}^T \right)^{-1} \boldsymbol{\gamma} \hat{\boldsymbol{\eta}} \xrightarrow[n \to \infty]{} \chi_2^2. \qquad (11)$$

The matrix $\hat{\Sigma}_{\hat{\boldsymbol{\eta}}}$ is obtained from the matrix $\hat{\Sigma}_{\hat{\boldsymbol{\theta}}}$ eliminating the rows and columns corresponding to $\hat{p}$, $\hat{\alpha}_1$ and $\hat{\alpha}_0$.

The global hypothesis global test (2) can also be solved from the individual hypothesis tests, i.e. $H_0 : PPV_1 = PPV_2$ and $H_0 : NPV_1 = NPV_2$, each one of them independently to an error $\alpha$, when the corresponding test statistics are

$$z = \frac{\widehat{PPV_1} - \widehat{PPV_2}}{\sqrt{\hat{V}ar\left(\widehat{PPV_1}\right) + \hat{V}ar\left(\widehat{PPV_2}\right) - 2Cov\left(\widehat{PPV_1}, \widehat{PPV_2}\right)}} \qquad (12)$$

and

$$z = \frac{\widehat{NPV_1} - \widehat{NPV_2}}{\sqrt{\hat{V}ar\left(\widehat{NPV_1}\right) + \hat{V}ar\left(\widehat{NPV_2}\right) - 2Cov\left(\widehat{NPV_1}, \widehat{NPV_2}\right)}}, \qquad (13)$$

both with normal standard distributions when the sample size is large.

Another method to solve the global test consists of solving each one of the individual tests along with a method of multiple comparisons, such as the classic method of Bonferroni [17] or the Holm method [18]. The Bonferroni method consists of solving each individual test to an error $\alpha/2$, and the Holm method is less conservative than the Bonferroni method.

## 3. Multiple imputation

Multiple imputation [19, 20, 21] is an alternative method to the *EM* algorithm which is used to solve problems with missing data. Multiple imputation (*MI*) consists of constructing *M* sets of complete data, with $M \geq 2$, obtained replacing the missing data with *M* sets imputed independently. From each complete dataset the parameters are estimated, thereby obtaining *M* estimators of each parameter. Then the *M* estimators of each parameter are combined properly to obtain a global estimator of each parameter and its variance. From these combined values it is possible to obtain confidence intervals for each parameter and also to solve the hypothesis test. In the context of the comparison of parameters of two *BDTs*, Harel and Zhou [12] studied the comparison of the sensitivities (specificities) of two *BDTs* in the presence of missing data *MAR* through confidence intervals applying *MI*.

We then study the simultaneous comparison of the predictive values applying multiple imputation. Firstly, the MICE Method is introduced and the hypothesis test is solved.

*3.1. MICE Method*

For the imputation of the missing data, we have applied the Multiple Imputation by Chained Equations (*MICE*), a method which is also known as Fully Conditional Specification or Sequential Regression Multivariate Imputation. The *MICE* method requires us to assume that the missing data are *MAR*, and it has the advantage of being a very flexible method which can be used with variables binary, ordinal or continuous variables. In the article by White et al [22] there is a more detailed explanation of the imputation of binary variables (which are the type of variables involved in the problem posed here) with the *MICE* Method. Therefore, in our situation we have three random binary variables: $T_1$, $T_2$ and *D*. For the variables $T_1$ and $T_2$ there are no missing data, since the two *BDTs* have been applied to all of the individuals in a sample. Nevertheless, the variable *D* is missing for a subset of individuals in the sample, since the disease status is unknown for these individuals. Firstly, all missing values are filled in at random. The variable $D$ is then regressed on the variables $T_1$ and $T_2$ through a logistic regression. The estimation is thus restricted to individuals with observed $T_1$ and $T_2$. The missing values in $D$ are then replaced by simulated draws from the posterior predictive distribution of $D$. This process is called a cycle, and to stabilize the results, this process is repeated a determined number of times, finally obtaining a set of imputed data. In the situation that we study here, from the $3 \times 4$ table (see Table 1: observed frequencies in the presence of partial verification) $M$ $2 \times 4$ tables are imputed (see Table 1: complete data), and from each one of these $M$ $2 \times 4$ tables, we calculate the estimators of the positive (negative) predictive values, of their differences, and their variances-covariances. As in the case of the EM-SEM algorithms, the comparison of the predictive values can be solved from the global hypothesis test or from the individual hypothesis tests, each one of them to an error $\alpha$ or applying a multiple comparison method.

*3.2. Global Test*

The solution of multiparametric hypothesis tests in problems with missing data applying *MI* has been the subject of several studies. Li et al [23] proposed a Wald statistic based on the *F*-distribution to solve a multidimensional hypothesis test, and Li et al [24] solved the hypothesis test combining the *p*-values (or in an equivalent way, the Wald statistics) obtained in the $M$ sets of imputed data from the same *F*-distribution. Meng and Rubin [25] solved the multidimensional hypothesis applying the likelihood ratio test.

In the situation studied here, let $\boldsymbol{\eta} = \left(PPV_1, NPV_1, PPV_2, NPV_2\right)^T$, $\hat{\boldsymbol{\eta}}^{(m)} = \left(\widehat{PPV}_1^{(m)}, \widehat{NPV}_1^{(m)}, \widehat{PPV}_2^{(m)}, \widehat{NPV}_2^{(m)}\right)^T$ be the estimator of $\boldsymbol{\eta}$ in the *m*th complete dataset ($m = 1,...,M$) obtained by applying the *MICE* Method and $\bar{\boldsymbol{\eta}} = \frac{1}{M}\sum_{m=1}^{M}\hat{\boldsymbol{\eta}}^{(m)}$ the overall estimate of $\boldsymbol{\eta}$. The global test (2) can be solved through the methods that are now described.

a) *Wald Test*

Let $\hat{\boldsymbol{\Sigma}}^{(m)}$ be the estimated variances-covariances matrix of $\hat{\boldsymbol{\eta}}^{(m)}$. The matrix $\hat{\boldsymbol{\Sigma}}^{(m)}$ is calculated from the *m*th dataset applying the delta method, and its elements are shown in Appendix *A*.

Let $\bar{\boldsymbol{\Sigma}} = \frac{1}{M}\sum_{m=1}^{M}\hat{\boldsymbol{\Sigma}}^{(m)}$, $\mathbf{B} = \frac{1}{M-1}\sum_{m=1}^{M}\left(\hat{\boldsymbol{\eta}}^{(m)} - \bar{\boldsymbol{\eta}}\right)\left(\hat{\boldsymbol{\eta}}^{(m)} - \bar{\boldsymbol{\eta}}\right)^{T}$ and $r_1 = \left(1 + M^{-1}\right)\text{trace}\left(\mathbf{B}\bar{\boldsymbol{\Sigma}}^{-1}\right)/2$.

The matrix $\bar{\boldsymbol{\Sigma}}$ measures the within imputation variability, the matrix $\mathbf{B}$ measures the between imputation variability and $r$ is the estimated average odds ratio of the fractions of missing information. Then, Wald test statistic [23] for the global test (10) is

$$F_1 = \frac{\bar{\boldsymbol{\eta}}^T \boldsymbol{\gamma}^T \left(\boldsymbol{\gamma} \bar{\boldsymbol{\Sigma}} \boldsymbol{\gamma}^T\right)^{-1} \boldsymbol{\gamma} \bar{\boldsymbol{\eta}}}{2(1+r_1)}, \quad (14)$$

whose distribution is one *F* with 2 (the dimension of the vector $\bar{\boldsymbol{\eta}}$) and

$$l = \begin{cases} 4 + (2M-6)\left[1 + \dfrac{M-2}{(M-1)r_1}\right]^2, & \text{if } 2(M-1) > 4 \\ \dfrac{3}{2}(M-1)\left(1 + r_1^{-1}\right)^2, & \text{if } 2(M-1) \leq 4 \end{cases}$$

degrees of freedom.

b) *Combination of p-values*

The solution of the global test can be made combining the *p*-values obtained in each one of the M complete datasets [24], or what amounts to the same thing, combining the Wald statistics. For the *m*th set of imputed data the Wald test statistic is

$$F^{(m)} = \left(\hat{\boldsymbol{\eta}}^{(m)}\right)^T \boldsymbol{\gamma}^T \left(\boldsymbol{\gamma}\hat{\boldsymbol{\Sigma}}^{(m)}\boldsymbol{\gamma}^T\right)^{-1} \boldsymbol{\gamma}\hat{\boldsymbol{\eta}}^{(m)}.$$

Let $\bar{F} = \dfrac{1}{M}\sum_{m=1}^{M} F^{(m)}$ and

$$r_2 = \left(1+M^{-1}\right)\left[\dfrac{1}{M-1}\sum_{m=1}^{M}\left(\sqrt{F^{(m)}} - \overline{\sqrt{F}}\right)^2\right],$$

where $\overline{\sqrt{F}} = \dfrac{1}{M}\sum_{m=1}^{M}\sqrt{F^{(m)}}$. The quantity $\bar{F}$ is the average Wald statistic and $r_2$ is $\left(1+M^{-1}\right)$ times the sample variance of $\left\{\sqrt{F^{(m)}},\ m=1,\ldots,M\right\}$. The statistic for the global test [25] is

$$F_2 = \dfrac{1}{1+r_2}\left(\dfrac{\bar{F}}{2} - \dfrac{M+1}{M-1}r_2\right), \tag{15}$$

when the combined p-value is

$$p-value = P\left(F_{2,l} \geq F_2\right) \tag{16}$$

with $l = 2^{-3/M}(M-1)\left(1+r_2^{-1}\right)^2$.

*c) Combined likelihood-ratio tests*

A third method to solve the global test is combining likelihood-ratio tests [25]. Let $\mathbf{z}^{(m)} = \left(x_{11}^{(m)}, x_{10}^{(m)}, x_{01}^{(m)}, x_{00}^{(m)}, y_{11}^{(m)}, y_{10}^{(m)}, y_{01}^{(m)}, y_{00}^{(m)}\right)^T$ be the vector of imputed frequencies in the $m$th complete dataset. The frequencies $x_{ij}^{(m)}$ $\left(y_{ij}^{(m)}\right)$ make reference to the individuals with the disease (without the disease) and with results $(i,j)$ of the diagnostic tests. Let $\phi_{ij}^{(m)}$ and $\varphi_{ij}^{(m)}$ be the probabilities corresponding to each cell of the imputed $2\times 4$ table, whose expressions are similar to those given in equations (7) adding the superindex ($m$) to all of the parameters. Let $\boldsymbol{\psi}^{(m)} = \left(\phi_{11}^{(m)}, \phi_{10}^{(m)}, \phi_{01}^{(m)}, \phi_{00}^{(m)}, \varphi_{11}^{(m)}, \varphi_{10}^{(m)}, \varphi_{01}^{(m)}, \varphi_{00}^{(m)}\right)^T$ be the corresponding vector of probabilities. The complete-data log-likelihood function is

$$L^{(m)}\left(\boldsymbol{\psi}^{(m)}; \mathbf{z}^{(m)}\right) \propto \sum_{i,j=0}^{1} x_{ij}^{(m)}\log\left[\phi_{ij}^{(m)}\right] + \sum_{i,j=0}^{1} y_{ij}^{(m)}\log\left[\varphi_{ij}^{(m)}\right].$$

Maximizing this function it holds that $\hat{\phi}_{ij}^{(m)} = x_{ij}^{(m)}/n$ and $\hat{\varphi}_{ij}^{(m)} = y_{ij}^{(m)}/n$, which are the non-restricted estimators of $\psi^{(m)}$ (i.e. $\hat{\psi}^{(m)}$), with $i,j = 0,1$. If in the $m$th set of imputed data the null hypothesis is true, i.e. $H_0 : \left( PPV_1^{(m)} = PPV_2^{(m)} \text{ and } NPV_1^{(m)} = PPV_2^{(m)} \right)$ is true, then the log-likelihood function is

$$L_0^{(m)}\left(\psi_0^{(m)}, \mathbf{z}^{(m)}\right) \propto x_{11}^{(m)} \log\left[\phi_{11}^{*(m)}\right] + \left(x_{10}^{(m)} + x_{01}^{(m)}\right)\log\left[\phi_{10}^{*(m)}\right] + x_{00}^{(m)} \log\left[\phi_{00}^{*(m)}\right]$$
$$+ y_{11}^{(m)} \log\left[\varphi_{11}^{*(m)}\right] + \left(y_{10}^{(m)} + y_{01}^{(m)}\right)\log\left[\varphi_{10}^{*(m)}\right] + y_{00}^{(m)} \log\left[\varphi_{00}^{*(m)}\right],$$

where $\phi_{ij}^{*(m)}$ and $\varphi_{ij}^{*(m)}$ are the probabilities subject to the null hypothesis. The demonstration can be seen in Appendix B. Maximizing this function it holds that

$$\hat{\phi}_{ij}^{*(m)} = \begin{cases} x_{ii}^{(m)}/n, & \text{if } i = j \\ \left[x_{10}^{(m)} + x_{01}^{(m)}\right]/(2n), & \text{if } i \neq j \end{cases}$$

and

$$\hat{\varphi}_{ij}^{*(m)} = \begin{cases} y_{ii}^{(m)}/n, & \text{if } i = j \\ \left[y_{10}^{(m)} + y_{01}^{(m)}\right]/(2n), & \text{if } i \neq j, \end{cases}$$

which are the estimators of $\psi^{(m)}$ subject to the null hypothesis (i.e. $\hat{\psi}_0^{(m)}$), with $i,j = 0,1$. Performing algebraic operations, the likelihood-ratio test statistic for the test $H_0 : \tau^{(m)} = 0$ and $\upsilon^{(m)} = 0$ vs $H_1 : \tau^{(m)} \neq 0$ and/or $\upsilon^{(m)} \neq 0$ is

$$F_3^{(m)} = 2\left[L^{(m)}\left(\hat{\psi}^{(m)}; \mathbf{z}^{(m)}\right) - L_0^{(m)}\left(\hat{\psi}_0^{(m)}; \mathbf{z}^{(m)}\right)\right] =$$
$$2\left[x_{10}^{(m)} \log\left(\frac{2x_{10}^{(m)}}{x_{10}^{(m)} + x_{01}^{(m)}}\right) + x_{01}^{(m)} \log\left(\frac{2x_{01}^{(m)}}{x_{10}^{(m)} + x_{01}^{(m)}}\right) + y_{10}^{(m)} \log\left(\frac{2y_{10}^{(m)}}{y_{10}^{(m)} + y_{01}^{(m)}}\right) + y_{01}^{(m)} \log\left(\frac{2y_{01}^{(m)}}{y_{10}^{(m)} + y_{01}^{(m)}}\right)\right].$$

Let $\bar{F}_3 = \frac{1}{M}\sum_{m=1}^{M} F_3^{(m)}$ be the average of likelihood-ratio statistics, and let $\bar{\psi} = \frac{1}{M}\sum_{m=1}^{M} \hat{\psi}^{(m)}$ and $\bar{\psi}_0 = \frac{1}{M}\sum_{m=1}^{M} \hat{\psi}_0^{(m)}$ be the vectors whose components are the measures of the estimators in the $M$ sets of imputed data subject to the non-restricted model and subject to the null hypothesis, respectively. Let

$$\tilde{F}_3 = \frac{2}{M}\sum_{m=1}^{M}\left[L^{(m)}\left(\overline{\psi};\mathbf{z}^{(m)}\right) - L_0^{(m)}\left(\overline{\psi}_0;\mathbf{z}^{(m)}\right)\right],$$

which is the measure of the likelihood-ratio statistics each one of which is assessed in $\overline{\psi}$ and $\overline{\psi}_0$, and let

$$r_3 = \frac{M+1}{2(M-1)}\left(\overline{F}_3 - \tilde{F}_3\right).$$

Finally, the likelihood-ratio test statistic for the global test is [25]

$$F_3 = \frac{\tilde{F}_3}{2(1+r_3)}, \qquad (17)$$

which is distributed according to a *F*-distribution with 2 and

$$l = \begin{cases} 4 + (2M-6)\left[1 + \dfrac{M-2}{(M-1)r_3}\right]^2, & \text{if } 2(M-1) > 4 \\ \dfrac{3}{2}(M-1)\left(1+r_3^{-1}\right)^2, & \text{if } 2(M-1) \leq 4 \end{cases}$$

degrees of freedom.

3.3. Individual Tests

As with the *EM* and *SEM* algorithms, the global test can also be solved from the individual tests applying multiple imputation with methods that compare the positive and negative predictive values independently or with a method of multiple comparisons. The methods that are going to be considered to individually compare the predictive values are those of Leisenring et al [3], Wang et al [4] and Kosinski [5]. For the method by Wang et al and for the method by Kosinski, the combination of results is achieved applying the rules of Rubin [19]. For the method by Leisenring et al, the combination of results is achieved calculating the average statistic. The test statistics of the method by Wang et al and of the method by Kosinski are of the type $\hat{\theta}/\sqrt{\hat{Var}(\hat{\theta})}$, and therefore the combination of results is achieved applying the rules of Rubin. Nevertheless, in the case of the method by Leisenring et al, the test statistic to compare the equality of the two positive (negative) predictive values is not of the type $\hat{\theta}/\sqrt{\hat{Var}(\hat{\theta})}$, and therefore the rules of Rubin cannot be applied. The methods of combination of the results are now described.

In each one of the $M$ sets of complete data, we estimate the positive (negative) predictive values and their variances-covariances. The positive predictive values are estimated as

$$\widehat{PPV}_1^{(m)} = \frac{x_{11}^{(m)} + x_{10}^{(m)}}{x_{11}^{(m)} + x_{10}^{(m)} + y_{11}^{(m)} + y_{10}^{(m)}} \quad \text{and} \quad \widehat{PPV}_2^{(m)} = \frac{x_{11}^{(m)} + x_{01}^{(m)}}{x_{11}^{(m)} + x_{01}^{(m)} + y_{11}^{(m)} + y_{01}^{(m)}}$$

and the negative predictive values as

$$\widehat{NPV}_1^{(m)} = \frac{y_{01}^{(m)} + y_{00}^{(m)}}{x_{01}^{(m)} + x_{00}^{(m)} + y_{01}^{(m)} + y_{00}^{(m)}} \quad \text{and} \quad \widehat{NPV}_2^{(m)} = \frac{y_{10}^{(m)} + y_{00}^{(m)}}{x_{10}^{(m)} + x_{00}^{(m)} + y_{10}^{(m)} + y_{00}^{(m)}}.$$

Appendix A shows the estimations of their variances-covariances. The results obtained are then combined applying the rules of Rubin. Firstly, we calculate the overall estimate of the difference between the two positive predictive values

$$\overline{PPV} = \frac{1}{M}\sum_{m=1}^{M}\left(\widehat{PPV}_1^{(m)} - \widehat{PPV}_2^{(m)}\right),$$

and their variance

$$\hat{Var}\left(\overline{PPV}\right) = \overline{Var}\left(\widehat{PPV}\right) + \frac{1}{M+1}B,$$

where

$$\overline{Var}\left(\widehat{PPV}\right) = \frac{1}{M}\sum_{m=1}^{M}\hat{Var}\left(\widehat{PPV}_1^{(m)} - \widehat{PPV}_2^{(m)}\right)$$

is the complete data variance estimate and

$$B = \frac{1}{M-1} \sum_{m=1}^{M} \left( \widehat{PPV}^{(m)} - \overline{PPV} \right)^2$$

is the between imputation variance. Finally, the test statistic for the test $H_0 : PPV_1 = PPV_2$ is

$$\frac{\overline{PPV}}{\sqrt{\hat{V}ar\left(\overline{PPV}\right)}}, \tag{18}$$

whose distribution is (Rubin, 1987) a *t*-distribution with $v = (M-1)\left(1 + \frac{M}{M+1} \frac{\hat{V}ar\left(\overline{PPV}\right)}{B}\right)$

degrees of freedom. The comparison of the negative predictive values is made in a similar way substituting *PPV* with *NPV*.

For the method by Leisenring et al, in the *m*th complete dataset the test statistic for $H_0 : PPV_1 = PPV_2$ is $z_{PPV}^{(m)}$ (its expression can be seen in Appendix A), whose distribution is a normal standard one when the sample size is large. Then for the Central Limit Theorem, the average of all of the test statistics $\overline{z}_{PPV} = \frac{1}{M} \sum_{m=1}^{M} z_{PPV}^{(m)}$ has a normal standard distribution when *M* is large. The process for the test $H_0 : NPV_1 = NPV_2$ is similar to the previous one.

## 4. Simulation experiments

Monte Carlo simulation experiments were carried out to study the Type I errors and the power of the hypothesis tests studied in Sections 2 and 3, as well as the relative biases of the estimators of the predictive values obtained with both methods. These experiments consisted of generating $N = 10000$ random samples of multinomial distributions sized $n = \{50, 100, 200, 500, 1000, 2000\}$, and whose probabilities were calculated from equations (5). As predictive values we considered the values $\{0.70, 0.75, 0.80, 0.85, 0.90, 0.95\}$, which are values that appear quite frequently in clinical practice, and as disease prevalence we took the values $p = \{25\%, 50\%, 75\%\}$. Once the predictive values and the disease prevalence, are set, the sensitivity and the specificity of each *BDT* were calculated from equations (3). As values of the dependence factors $\alpha_i$ we considered intermediate and high values. Finally, the probabilities of the multinomial distributions were calculated applying equations (5). Therefore, the probabilities of the multinomial distributions were calculated from the predictive values, and we did not set *a priori* the values of the sensitivities and the specificities of the *BDTs*. The simulation experiments were designed in such a way that in all of the samples generated it is possible to apply the *EM-SEM* algorithms and multiple imputation. Therefore, if in a sample a frequency $a_{ij}$ (or $b_{ij}$) is zero then it is not possible to apply multiple imputation (it is not possible to apply logistic regression to impute the missing data), then this sample was discarded and another one was generated instead until completing the *N* samples. Regarding the *EM* and *SEM* algorithms, we established as a stop criterion $\delta = 10^{-12}$ and $\sqrt{\delta} = 10^{-6}$ respectively. Regarding multiple imputation, for each one of the *N* random samples $M = 20$ compete data sets were generated and 100 cycles were performed. In the first phase simulations were made considering $M = 20$ and $M = 50$ and performing 100 and 200 cycles in each case, obtaining very similar results; therefore, to reduce computation time we finally considered $M = 20$ and 100 cycles. For all of the study, we set as the nominal error $\alpha = 5\%$, and considered that a method overwhelms the nominal error or exceeds it too much when its Type I error is higher than 7%. The simulation experiments were carried out with R [26] and for the multiple imputation we used the library "mice" [27].

Therefore, in the simulation experiments we studied and compared the Type I errors and the power of sixteen different methods to solve the global hypothesis test

$$H_0 : (PPV_1 = PPV_2 \text{ and } NPV_1 = NPV_2) \text{ vs } H_1 : (PPV_1 \neq PPV_2 \text{ and/or } NPV_1 \neq NPV_2).$$

Of the sixteen methods, four are based on the *EM* and *SEM* algorithms and the other twelve on multiple imputation. The methods based on the *EM* and *SEM* algorithms are: (1) a global hypothesis global test based on the chi-squared distribution with $\alpha = 5\%$; (2) an individual comparison of the positive and negative predictive values with $\alpha = 5\%$, Bonferroni and Holm. The twelve methods based on multiple imputation are: (1) a global hypothesis test applying the Wald method, the combination of *p*-values and the combined likelihood ratio tests, all of them with $\alpha = 5\%$; (2) an individual comparison of the positive and negative predictive values applying the methods of Leisenring et al, Wang et al and Kosinski, each of them with $\alpha = 5\%$, Bonferroni and Holm.

*4.1. EM and SEM Algorithms*

Table 2 shows some of the results obtained for the Type I errors of the different methods applying the *EM* and *SEM* algorithms. In this Table, Method 1 consists of comparing the predictive values solving the individual hypothesis tests each of them to an error $\alpha = 5\%$, and Method 2 consists of comparing the predictive values solving the individual hypothesis tests along with the Bonferroni method to an error $\alpha = 5\%$. The results obtained applying the Holm method to an error $\alpha = 5\%$ are not shown as they are practically the same as those obtained with the Bonferroni method. In general terms, the Type I errors of all of the methods increase when the verification probabilities increase, and decrease when the covariances $\alpha_i$ increase. In general terms, all of the methods are very conservative when the sample size is small $(n = 50)$ or moderate $(n = 100 - 200)$. When the sample size is large $(n \geq 500)$, depending on the verification probabilities and on the covariances $\alpha_i$, the global test has a Type I error that fluctuates around the nominal error. On some occasions, above all when the verification probabilities are low or the covariances are high, the Type I error of the global test may slightly exceed the nominal error without actually overwhelming it. Method 1 may overwhelm the nominal error, above all when the sample size is large. Method 2 has a Type I error whose behaviour is very similar to that of the global test.

Regarding the power of these methods, Table 3 shows some results. The power of these methods increases when there is an increase in in the verification probabilities, whereas the covariances $\alpha_i$ do not have a clear effect upon the power. All of the methods have a very small power when the sample size is small $(n=50)$ or moderate $(n=100-200)$, and it is necessary to have a large sample size (depending on the verification probabilities) so that the power is higher (over 80%). Although there are no clear rules, in general terms the global test is normally more powerful than Method 2. Regarding Method 1, there are also no clear rules. Sometimes the global test is more powerful and on other occasions Method 1 is more powerful (it is a method that easily overwhelms the nominal error), depending on the values that the predictive values take.

From the results of the simulation experiments applying the *EM* and *SEM* algorithms, the method to compare the predictive values of the two *BDTs* with the best asymptotic behaviour is the global test, since its Type I error does not exceed the nominal error too much and, in general terms, it has more power than Method 1 (this is a method whose Type I error also does not exceed the nominal error too much). Method 2 may exceed the nominal error too much and, therefore, lead to false significances.

*4.2. Multiple Imputation*

Table 4 shows the results through multiple imputation for the same scenarios as Table 2. In this table, method *L*1 refers to the individual comparison of the predictive values applying the method of Leisenring et al with $\alpha = 5\%$, and method *L*2 refers to the individual comparison of predictive values applying the method of Leisenring et al along with the Bonferroni method with $\alpha = 5\%$. The methods *W*1 and *W*2 (*K*1 and *K*2) are the method by Wang et al (Kosinski) with $\alpha = 5\%$ and Bonferroni respectively. The results obtained applying the Holm method are not shown as they are practically identical those obtained with Bonferroni. As with the *EM* and *SEM* algorithms, the Type I errors of all the methods based on multiple imputation increase when the verification probabilities increase, and decrease when the covariances $\alpha_i$ increase.

Regarding the global tests, the Type I error of the test based on the combination of *p*-values is very similar to the Type I error of the combined likelihood-ratio tests, both of which fluctuate around the nominal error when the sample size is large. The global test based on the Wald test is very conservative (even when the sample size is large), and its Type I error is smaller than that of the other two methods.

The methods based on the individual comparisons to an error $\alpha = 5\%$ (methods *L*1, *W*1 and *K*1) have Type I errors that may exceed the nominal error too much, above all when the sample size is large. Therefore, these methods may lead to an excess of false significances. Regarding the methods based on the individual tests along with Bonferroni, method *L*2 has a Type I error which may exceed the nominal error too much when the sample size is large. Method *K*2 has a Type I error with better fluctuations around the nominal error (with exceeding it too much) than method *W*2, above all when the sample size is large. In general terms, there is no important difference between the Type I error of method *K*2 and the Type I error of the global test based on the combination of *p*-values (or combined likelihood-ratio tests), above all when the sample size is large.

Regarding the power, Table 5 shows the results for the same scenarios given in Table 3. The power of all of the methods increases when the verification probabilities increase, whereas the covariances $\alpha_i$ do not have a clear effect upon the power.

In general terms, the global test based on the combination of *p*-values is more powerful than the combined likelihood-ratio tests, and the difference is greater when the verification probabilities are low than when they are high. Moreover, both methods are more powerful tan the Wald test (as this test is very conservative in relation to the other two).

Comparing the global test based on the combination of *p*-values in relation to the Methods *L*1, *W*1 and *K*1 (based on the individual comparisons to an $\alpha = 5\%$ ), there are no clear rules about their behaviour. Sometimes the global test is more powerful and on other occasions these methods are more powerful (methods which may clearly overwhelm the nominal error), depending on the verification probabilities and on the values that the predictive values take.

Regarding Method *L*2, in general terms the global test based on the combination of *p*-values is less powerful, due to the fact that the Type I error of Method *L*2 (which may clearly overwhelm the nominal error) is greater than that of the global test (which does not overwhelm the nominal error). Regarding methods *W*2 and *K*2, there is no important difference between their powers, and these powers are a little lower than those of the global test based on the combination of *p*-values, above all when the sample size is large.

Having analysed the results of the simulation experiments applying multiple imputation, the method to compare the predictive values of two *BDTs* with the best asymptotic behaviour is the global test based on the combination of *p*-values, since its Type I error does not overwhelm the nominal error and its power is somewhat higher than that of the other methods which do not overwhelm the nominal error.

*4.3. Relative Biases*

Table 6 shows some results for the relative biases of the estimators of the predictive values applying the *EM* algorithm and applying multiple imputation. The relative biases decrease when the verification probabilities Increase whereas the covariances $\alpha_i$ have practically no effect on the estimators obtained applying both methods. In general terms, the difference between the relative biases obtained with both methods is very small, and therefore both methods lead to estimations which on average are very similar to each other.

*4.4. EM and SEM algorithms or multiple imputation?*

Comparing the Type I error of the global test based on the *EM-SEM* algorithms (Table 2) with the Type I error of the global test based on the combination of *p*-values (Table 4), In general terms, there is no important difference among them. Regarding the power of both methods, in general terms the power of the test based on multiple imputation (combination of *p*-values) is a little higher than the power of the test based on the *EM* and *SEM* algorithms when the sample is small or moderate; when the sample is large, the power of both methods is very similar. Therefore, from the simulation experiments, it is possible to give the following general rule of application based on the sample size:

    a) Apply the *EM-SEM* algorithms or multiple imputation when the sample size is large.

    b) Apply multiple imputation based on the combination of *p*-values when the sample size is small or moderate.

*4.5. Causes of the significance*

When the *EM* and *SEM* algorithms are applied, if the global test is not significant to an error $\alpha$ then we do not reject the homogeneity of the predictive values of both *BDTs*. If the global test is significant to an error $\alpha$, then the causes of the significance are investigated solving the individual hypothesis tests, i.e. $H_0: PPV_1 = PPV_2$ and $H_0: NPV_1 = NPV_2$, along with the Bonferroni (or Holm) method to an error $\alpha$. The application of the individual tests along with the Bonferroni (or Holm) method is justified, just as in the simulation experiments, by the fact that this method has a Type I error which does not overwhelm the nominal error.

When multiple imputation is applied, the investigation of the causes of the significance is made in a similar way to in the previous case. We solve the global test based on the combination of *p*-values to an error $\alpha$ and if the test is not significant then we do not reject the homogeneity of the predictive values. If the global test is significant, then the causes of the investigation are studied solving the individual tests applying the Kosinski method along with the Bonferroni (or Holm) method to an error $\alpha$ for any sample size, although it is also possible to apply the method of Wang et al along with Bonferroni (or Holm) when the sample size is small or moderate. When the sample size is large, the Kosinski method has a better Type I error behaviour than the method of Wang et al, and their power is very similar. When the sample size is small or moderate, the Kosinski method and the method of Wang et al have very similar Type I errors and powers.

## 5 EmSemPv and ImPv Programmes

Two programmes in R were written: EmSemPv and ImPv. The EmSemPv programme compares the predictive values of two diagnostic tests in the presence of partial disease verification applying the *EM-SEM* algorithms and the ImPv programme solves the same problem applying multiple imputation (*MICE* method). Both programmes are available as supplementary material for this article.

The EmSemPv programme is run with the command

$$\text{emsempv}(a_{11}, a_{10}, a_{01}, a_{00}, b_{11}, b_{10}, b_{01}, b_{00}, c_{11}, c_{10}, c_{01}, c_{00}).$$

The programme checks that the values of the frequencies are feasible (for example, that there are no negative values, frequencies with decimals, etc…). By default, the stop criterion of the *EM* algorithm is $10^{-12}$, the confidence for the calculation of the intervals is 95% and, as initial values of the *EM* algorithm, the values $d_{ij}^{(0)} = c_{ij}/2$ are used. The programme provides the estimations of the predictive values and their corresponding standard errors, the inverse Fisher information matrix of the complete data, the *DM* matrix, the estimated variances-covariances matrix of the predictive values, the test statistics and p-values of the global test and of the individual tests, as well as the confidence intervals for the difference in the positive (negative) predictive values.

The ImPv programme solves the problem applying the *MICE* method and is run with the command

$$\mathrm{impv}(a_{11}, a_{10}, a_{01}, a_{00}, b_{11}, b_{10}, b_{01}, b_{00}, c_{11}, c_{10}, c_{01}, c_{00}).$$

The programme checks that the values of the frequencies are feasible (for example, that there is no frequency equal to zero, negative values, frequencies with decimals, etc…). By default, 20 complete datasets are generated, for each complete dataset 100 cycles are performed and the confidence for the calculation of the intervals is 95%. The programme provides the estimations of the predictive values and their corresponding standard errors, the estimated variances-covariances matrix of the predictive values, the test statistics and *p*-values of the global test and of the individual tests applying the method of Wang et al and the Kosinski method, as well as the confidence intervals for the difference in the positive (negative) predictive values.

## 6. Example

The results obtained were applied to the study by Hall et al [28] on the diagnosis of Alzheimer's disease. Hall et al used two diagnostic tests for the diagnosis of Alzheimer's disease: a new diagnostic test(*Test* 1) based on a cognitive test applied to the patient and in another relative test to another person who knows the patient, and a standard diagnostic test based on a cognitive test (*Test* 2), and as the *GS* they used a clinical assessment (neurological examination, computerized tomography, neuropsychological and laboratory tests…). Table 7 shows the results obtained applying the two diagnostic tests to a sample of 588 patients over 75 years of age, and where the random variable $T_1$ models the result of *Test* 1, the random variable $T_2$ models the result of Test 2 and the random variable $D$ models the result of the *GS*. The study by Hall et al corresponds to a two-phase study: in the first phase, the two diagnostic tests were applied to all of the patients, and in the second phase the *GS* was only applied to a subset of patients depending on the results of both diagnostic tests. Consequently, it is assumed that the verification process is *MAR*. As the sample size is large, the comparison of the predictive values of both *BDTs* can be made applying the *EM-SEM* algorithms and multiple imputation.

Table 7 (results with *EM-SEM* algorithms) shows the estimations (with three decimal figures) of the predictive values and their standard errors (*SE*) obtained applying the EmSemPv programme with the command "emsempv (31,5,3,1,25,10,19,55,22,6,65,346)". The *EM* algorithm has converged in 186 iterations. The inverse Fisher information matrix of the complete data, the *DM* matrix and the variances-covariances matrix can be seen when the programme is run. The test statistic for the global test is $Q^2 = 30.097$ and the *p*-value is $2.914 \times 10^{-7}$, and therefore with an error $\alpha = 5\%$ we reject the equality of the predictive values of both *BDTs*. Solving the individual tests $H_0 : PPV_1 = PPV_2$ and $H_0 : NPV_1 = NPV_2$ it holds that the respective test statistics are $z = 3.251$ ($p$-value $= 0.001$) and $z = 0.362$ ($p$-value $= 0.718$). Applying the Bonferroni (or Holm) method to an error $\alpha = 5\%$ we reject the equality of the two positive predictive values and we do not reject the equality of the two negative predictive values. When the two diagnostic are applied to the population being studied, the positive predictive value of *Test* 1 is significantly higher than that of *Test* 2 (95% confidence interval for the difference: 0.069 to 0.278).

Table 7 (results with the *MICE* method) shows the estimations obtained applying the ImPv programme with the command "impv (31,5,3,1,25,10,19,55,22,6,65,346)". The test statistic for the global test applying the method of the combination of *p*-values is $F_2 = 15.974$ and the *p*-value is $1.291 \times 10^{-7}$, and therefore with an error $\alpha = 5\%$ we reject the equality of the predictive values of both *BDTs*. Solving the individual tests $H_0 : PPV_1 = PPV_2$ and $H_0 : NPV_1 = NPV_2$ applying the Kosinski method it holds that the respective test statistics are $z = 4.808$ ($p$-value $= 1.527 \times 10^{-7}$) and $z = 0.747$ ($p$-value $= 0.455$). Applying the Bonferroni (or Holm) method to an error $\alpha = 5\%$ we reject the equality of the two positive (negative) predictive values and we do not reject the equality of the two negative predictive values. When the two diagnostic tests are applied to the population being studied, the positive predictive value of *Test* 1 is significantly greater than that of *Test* 2 (95% confidence interval for the difference: 0.101 to 0.239).

In this example, it can be observed how with both methods, *EM* and *SEM* algorithms and multiple imputation, the results obtained are very similar, both in terms of the point estimators and of their variances-covariances. Moreover, the conclusions are the same: we reject the equality between the two positive predictive values and we do not reject the equality between the two negative predictive values.

## 7. Discussion

The comparison of the predictive values of two *BDTs* is a topic of interest in the study of the statistical methods for medical diagnosis. This article studies the computational methods to solve this problem in the presence of missing data. The comparison of the two positive predictive values and of the two negative predictive values was studied simultaneously applying the *EM-SEM* algorithms and multiple imputation. With both methods it is required that the missing data be *MAR*, and therefore if the verification process depends on the disease status then this assumption is not verified and the methods cannot be applied.

Simulation experiments were carried out to study the asymptotic behaviour of the global test of comparison to study the asymptotic behaviour of the global test of comparison of the predictive values, and of other alternative methods, both with the *EM-SEM* algorithms and with multiple imputation, giving some general rules of application based on the sample size. In general terms, multiple imputation can be applied to any sample size, whereas the *EM-SEM* algorithms require the sample size to be large. If the global test is not significant to an $\alpha$ error then we do not reject the homogeneity of the predictive values of both *BDTs*. If the global test is significant then the causes of the significance are investigated comparing the two positive predictive values and the two negative predictive values independently and a method of multiple comparisons is applied, such as Bonferroni or Holm, which are very easy methods to apply. This procedure is very similar to a variance analysis: the global test is solved and if this is significant paired comparisons are carried out and a method of multiple comparisons is applied.

The application of the *EM-SEM* algorithms leads to the same results as the application of the Marín-Jiménez and Roldán-Nofuentes method [11]. These authors simultaneously compared the predictive values of multiple *BDTs* obtaining the estimators through the maximum likelihood method and estimating the variances-covariances applying the delta method. The advantage of the *EM-SEM* algorithms over the maximum likelihood method is that the former can be applied when some $s_{ij}$ or $r_{ij}$ frequency is equal to zero, whereas with the maximum likelihood method if any one these frequencies is equal to zero then it is not possible to estimate the variances-covariances.

As for multiple imputation, this can only be applied when all of the $a_{ij}$ and $b_{ij}$ frequencies are higher than zero, since in that case it is not possible to generate complete datasets through logistic regression. Furthermore, the global tests based on the Wald method and on the combination of *p*-values requires the fractions of missing information for all components of the parameter vector to be equal. When there are important differences between the fractions of missing information and these are large, there may be an important effect on the size of the test and on the power [23, 24]. Traditionally, Rubin [19] recommended imputing five complete datasets in order to be able to apply multiple imputation. In the situation analysed in this article, in the initial simulation experiments $M = \{20, 50\}$ compete datasets were considered, and the results were very similar, and therefore it was decided to generate $M = 20$ complete datasets to save computation time. The simulation experiments demonstrated that the global test based on the combination of *p*-values (which is the test based on multiple imputation with the best asymptotic behaviour) has a Type I error which is very similar to the global test based on the *EM-SEM* algorithms. Regarding power, that of the global test based on the combination of *p*-values is a little higher than that of the global test based on the *EM-SEM* algorithms when the sample is small or moderate, and they are very similar when the sample is large. Therefore, the number of complete datasets was sufficiently large, and did not have any negative effect on the size and the power of the global test.

**Disclosure statement**

No potential conflict of interest was reported by the authors.

**Funding**

This research was supported by the Spanish Ministry of Economy [Grant Number MTM2016-76938-P].

# Appendix A

Let us consider that two *BDTs* are applied to all the individuals in a random sample sized *n*, whose disease status (present or absent) is known through the application of a *GS*. Let $x_{ij}$ $(y_{ij})$ be the numbers of diseased (non-diseased) individuals among whom *Test* 1 leads to a result *i* and *Test* 2 leads to result *j*, with $i,j = 0,1$ (0 indicates a negative result and 1 a positive result). We now summarize the methods of Leisenring et al [3], Wang et al [4] and Kosinski [5]. The Tsou method [6] is not considered since it is equivalent to the Kosinski method.

*Method of Leisenring et al*

Leisenring et al studied the comparison of the positive and negative predictive values of two binary tests through marginal regression models, and they were able to estimate these models separately or jointly using *GEE* models. Leisenring et al deduced score statistics to compare the positive and negative predictive values of two binary tests in paired designs. Using the notation from the previous Section, the score statistic for the test $H_0: PPV_1 = PPV_2$ is

$$z_{PPV} = \frac{x_{11}(1-2\bar{Z}_1) + x_{01}(1-\bar{Z}_1) - x_{10}\bar{Z}_1}{\sqrt{x_{11}(1-\bar{D}_1)^2(1-2\bar{Z}_1)^2 + x_{01}(1-\bar{D}_1)^2(1-\bar{Z}_1)^2 + x_{10}(1-\bar{D}_1)^2\bar{Z}_1^2 + y_{11}\bar{D}_1^2(1-2\bar{Z}_1)^2 + y_{01}\bar{D}_1^2(1-\bar{Z}_1)^2 + y_{10}\bar{D}_1^2\bar{Z}_1^2}}$$

and the score statistic to compare the test $H_0: NPV_1 = NPV_2$ is

$$z_{NPV} = \frac{y_{00}(1-2\bar{Z}_2) + y_{10}(1-\bar{Z}_2) - y_{01}\bar{Z}_2}{\sqrt{y_{00}(1-\bar{D}_2)^2(1-2\bar{Z}_2)^2 + y_{10}(1-\bar{D}_2)^2(1-\bar{Z}_2)^2 + y_{01}(1-\bar{D}_2)^2\bar{Z}_2^2 + x_{00}\bar{D}_2^2(1-2\bar{Z}_2)^2 + x_{10}\bar{D}_2^2(1-\bar{Z}_2)^2 + x_{01}\bar{D}_2^2\bar{Z}_2^2}}.$$

Score statistics have a normal distribution when the null hypothesis is true, and where

$$\bar{Z}_1 = \frac{x_{11} + x_{01} + y_{11} + y_{01}}{2x_{11} + x_{01} + x_{10} + 2y_{11} + y_{10} + y_{01}}, \quad \bar{D}_1 = \frac{2x_{11} + x_{01} + x_{10}}{2x_{11} + x_{01} + x_{10} + 2y_{11} + y_{10} + y_{01}}.$$

$$\bar{Z}_2 = \frac{x_{00} + x_{10} + y_{00} + y_{10}}{2x_{00} + x_{01} + x_{10} + 2y_{00} + y_{01} + y_{10}} \quad \text{and} \quad \bar{D}_2 = \frac{2y_{00} + y_{01} + y_{10}}{2x_{00} + x_{01} + x_{10} + 2y_{00} + y_{01} + y_{10}}.$$

*Method of Wang et al*

Wang et al studied the comparison of the predictive values of two binary tests through a weighted least square method and compared their method to that of Leisenring et al, before recommending the comparison of the predictive values using the weighted least square method based on the difference between the two positive (negative) predictive values. The test statistics for $H_0: PPV_1 = PPV_2$ and $H_0: NPV_1 = NPV_2$ are

$$z_{PPV} = \frac{\widehat{PPV_1} - \widehat{PPV_2}}{\sqrt{\hat{V}ar(\widehat{PPV_1}) + \hat{V}ar(\widehat{PPV_1}) - 2\hat{C}ov(\widehat{PPV_1}, \widehat{PPV_2})}}$$

and

$$z_{NPV} = \frac{\widehat{NPV_1} - \widehat{NPV_2}}{\sqrt{\hat{V}ar(\widehat{NPV_1}) + \hat{V}ar(\widehat{NPV_1}) - 2\hat{C}ov(\widehat{NPV_1}, \widehat{NPV_2})}}$$

respectively, where

$$\widehat{PPV_1} = \frac{x_{11} + x_{10}}{x_{11} + x_{10} + y_{11} + y_{10}}, \quad \widehat{PPV_2} = \frac{x_{11} + x_{01}}{x_{11} + x_{01} + y_{11} + y_{01}}, \quad \widehat{NPV_1} = \frac{y_{01} + y_{00}}{x_{01} + x_{00} + y_{01} + y_{00}} \text{ and }$$

$$\widehat{NPV_2} = \frac{y_{10} + y_{00}}{x_{10} + x_{00} + y_{10} + y_{00}}.$$

Both test statistics follow a standard normal distribution, and the variances are estimated by applying the delta method (the expressions are shown in the method of Roldán-Nofuentes et al [7] which will now be summarized).

*Kosinski Method*

Kosinski proposed a weighted generalized score statistic to solve the hypothesis test of comparison of the predictive values. The weighted generalized score statistic for the test $H_0: PPV_1 = PPV_2$ is

$$z_\tau = \frac{\widehat{PPV_1} - \widehat{PPV_2}}{\sqrt{\left\{\widehat{PPV}_p\left(1-\widehat{PPV}_p\right) - 2C_p^{PPV}\right\}\left(\frac{1}{n_{10}+n_{11}} + \frac{1}{n_{01}+n_{11}}\right)}},$$

and the weighted generalized score statistic for the test $H_0: NPV_1 = NPV_2$ is

$$z_\upsilon = \frac{\widehat{NPV_1} - \widehat{NPV_2}}{\sqrt{\left\{\widehat{NPV}_p\left(1-\widehat{NPV}_p\right) - 2C_p^{NPV}\right\}\left(\frac{1}{n_{00}+n_{01}} + \frac{1}{n_{00}+n_{10}}\right)}},$$

which has a standard normal distribution when the null hypothesis is true, where

$$\widehat{PPV}_p = \frac{2x_{11} + x_{10} + x_{01}}{2n_{11} + n_{10} + n_{01}} \text{ and } \widehat{NPV}_p = \frac{2y_{00} + y_{01} + y_{10}}{2n_{00} + n_{01} + n_{10}}$$

are the pooled positive predictive value and pooled negative predictive value respectively, and

$$C_p^{PPV} = \frac{x_{11}\left(1-\widehat{PPV}_p\right)^2 + y_{11}\widehat{PPV}_p^{\ 2}}{2n_{11}+n_{10}+n_{01}}, \quad C_p^{NPV} = \frac{x_{00}\widehat{NPV}_p^{\ 2} + y_{00}\left(1-\widehat{NPV}_p\right)^2}{2n_{00}+n_{01}+n_{10}}$$

and $n_{ij} = x_{ij} + y_{ij}$.

*Method of Roldán-Nofuentes et al*

Roldán-Nofuentes et al [7] studied the simultaneous comparison of the predictive values of two *BDTs* subject to a paired design. The simultaneous comparison of the predictive values of two binary tests consists of solving the hypothesis test

$$H_0: (PPV_1 = PPV_2 \text{ and } NPV_1 = NPV_2) \text{ vs } H_1: (PPV_1 \neq PPV_2 \text{ and/or } NPV_1 \neq NPV_2),$$

Applying the delta method, the estimated variances-covariances of the estimators of the predictive values are:

$$\hat{V}ar\left(\widehat{PPV_1}\right) = \frac{(x_{10}+x_{11})(y_{10}+y_{11})}{n(x_{10}+x_{11}+y_{10}+y_{11})^3}, \quad \hat{V}ar\left(\widehat{NPV_1}\right) = \frac{(x_{00}+x_{01})(y_{00}+y_{01})}{n(x_{00}+x_{01}+y_{00}+y_{01})^3}$$

$$\hat{V}ar\left(\widehat{PPV_2}\right) = \frac{(x_{01}+x_{11})(y_{01}+y_{11})}{n(x_{01}+x_{11}+y_{01}+y_{11})^3}, \quad \hat{V}ar\left(\widehat{NPV_2}\right) = \frac{(x_{00}+x_{10})(y_{00}+y_{10})}{n(x_{00}+x_{10}+y_{00}+y_{10})^3},$$

$$\hat{C}ov\left(\widehat{PPV_1},\widehat{PPV_2}\right) = \frac{x_{01}x_{10}y_{11} + x_{11}\left[y_{01}(y_{10}+y_{11}) + y_{11}(x_{01}+x_{10}+x_{11}+y_{10}+y_{11})\right]}{(x_{01}+x_{11}+y_{01}+y_{11})^2(x_{10}+s_{11}+y_{10}+y_{11})^2},$$

$$\hat{C}ov\left(\widehat{PPV_1},\widehat{NPV_2}\right) = -\frac{x_{00}(x_{10}+x_{11})y_{10} + x_{10}y_{10}(x_{10}+x_{11}+y_{00}+r_{10}) + x_{10}(y_{00}+r_{10})y_{11}}{(x_{00}+x_{10}+y_{00}+y_{10})^2(x_{10}+x_{11}+y_{10}+y_{11})^2},$$

$$\hat{C}ov\left(\widehat{PPV_2},\widehat{NPV_1}\right) = -\frac{x_{00}(x_{01}+x_{11})y_{01} + x_{01}y_{01}(x_{01}+x_{11}+y_{00}+y_{01}) + x_{01}(y_{00}+y_{01})y_{11}}{(x_{00}+x_{01}+y_{00}+y_{01})^2(x_{01}+x_{11}+y_{01}+y_{11})^2},$$

$$\hat{C}ov\left(\widehat{NPV_1},\widehat{NPV_2}\right) = \frac{x_{00}(y_{00}+y_{01})y_{10} + y_{00}\left[y_{00}^2 + x_{01}x_{10} + x_{00}(x_{01}+x_{10}+y_{00}+y_{01})\right]}{(x_{00}+x_{01}+y_{00}+y_{01})^2(x_{00}+x_{10}+y_{00}+y_{10})^2},$$

$$\hat{C}ov\left(\widehat{PPV_1},\widehat{NPV_1}\right) = \hat{C}ov\left(\widehat{PPV_2},\widehat{NPV_2}\right) = 0.$$

The test statistic for $H_0:(PPV_1 = PPV_2$ and $NPV_1 = NPV_2)$ is $Q^2 = \hat{\boldsymbol{\eta}}^T\boldsymbol{\gamma}^T\left(\boldsymbol{\gamma}\hat{\Sigma}\boldsymbol{\gamma}^T\right)^{-1}\boldsymbol{\gamma}\hat{\boldsymbol{\eta}}$, where $\hat{\boldsymbol{\eta}} = \left(\widehat{PPV_1},\widehat{NPV_1},\widehat{PPV_2},\widehat{NPV_2}\right)^T$, $\hat{\Sigma}$ is the estimated variance-covariance matrix of $\hat{\boldsymbol{\eta}}$ and $\boldsymbol{\gamma}$ is the design matrix, i.e.

$$\boldsymbol{\gamma} = \begin{pmatrix} 1 & 0 & -1 & 0 \\ 0 & 1 & 0 & -1 \end{pmatrix}.$$

The test statistic $Q^2$ is distributed asymptotically according to a central chi-square distribution with two degrees of freedom if $H_0$ is true.

When all of these methods are used applying multiple imputation, all the equations are valid for the $m$th complete dataset, adding superindex $(m)$ to all of the terms of the equations

Table 1. Observed frequencies in the presence of partial verification and complete data.

| | Observed frequencies in the presence of partial verification | | | | |
|---|---|---|---|---|---|
| | $T_1 = 1$ | | $T_1 = 0$ | | |
| | $T_2 = 1$ | $T_2 = 0$ | $T_2 = 1$ | $T_2 = 0$ | Total |
| $V = 1$ | | | | | |
| $D = 1$ | $a_{11}$ | $a_{10}$ | $a_{01}$ | $a_{00}$ | $a$ |
| $D = 0$ | $b_{11}$ | $b_{10}$ | $b_{01}$ | $b_{00}$ | $b$ |
| $V = 0$ | $c_{11}$ | $c_{10}$ | $c_{01}$ | $c_{00}$ | $c$ |
| Total | $n_{11}$ | $n_{10}$ | $n_{01}$ | $n_{00}$ | $n$ |
| | Complete data | | | | |
| | $T_1 = 1$ | | $T_1 = 0$ | | |
| | $T_2 = 1$ | $T_2 = 0$ | $T_2 = 1$ | $T_2 = 0$ | Total |
| $D = 1$ | $a_{11} + d_{11}$ | $a_{10} + d_{10}$ | $a_{01} + d_{01}$ | $a_{00} + +d_{00}$ | $a + d$ |
| $D = 0$ | $b_{11} + c_{11} - d_{11}$ | $b_{10} + c_{10} - d_{10}$ | $b_{01} + c_{01} - d_{01}$ | $b_{00} + c_{00} - d_{00}$ | $b + c - d$ |
| Total | $n_{11}$ | $n_{10}$ | $n_{01}$ | $n_{00}$ | $n$ |

Table 2. Type I errors obtained by applying the *EM-SEM* algorithms.

| | $PPV_1 = PPV_2 = 0.85$ $NPV_1 = NPV_2 = 0.95$ | | | | | |
|---|---|---|---|---|---|---|
| | $Se_1 = Se_2 = 0.85$ $Sp_1 = Sp_2 = 0.95$ $p = 25\%$ | | | | | |
| | $\lambda_{11} = 0.50$ $\lambda_{10} = 0.30$ $\lambda_{01} = 0.30$ $\lambda_{00} = 0.05$ | | | | | |
| | $\alpha_1 = 1.09$ $\alpha_0 = 10.50$ | | | $\alpha_1 = 1.13$ $\alpha_0 = 15.25$ | | |
| $n$ | Global test | Method 1 | Method 2 | Global test | Method 1 | Method 2 |
| 50   | 0     | 0     | 0     | 0     | 0     | 0     |
| 100  | 0.001 | 0.003 | 0.001 | 0     | 0     | 0     |
| 200  | 0.002 | 0.006 | 0.001 | 0.001 | 0.003 | 0.001 |
| 500  | 0.012 | 0.036 | 0.013 | 0.007 | 0.014 | 0.002 |
| 1000 | 0.041 | 0.071 | 0.041 | 0.018 | 0.047 | 0.022 |
| 2000 | 0.046 | 0.073 | 0.042 | 0.034 | 0.052 | 0.021 |
| | $\lambda_{11} = 0.95$ $\lambda_{10} = 0.75$ $\lambda_{01} = 0.75$ $\lambda_{00} = 0.30$ | | | | | |
| | $\alpha_1 = 1.09$ $\alpha_0 = 10.50$ | | | $\alpha_1 = 1.13$ $\alpha_0 = 15.25$ | | |
| $n$ | Global test | Method 1 | Method 2 | Global test | Method 1 | Method 2 |
| 50   | 0     | 0     | 0     | 0     | 0     | 0     |
| 100  | 0.002 | 0.014 | 0.001 | 0     | 0.002 | 0     |
| 200  | 0.010 | 0.017 | 0.009 | 0.001 | 0.003 | 0.001 |
| 500  | 0.033 | 0.067 | 0.034 | 0.011 | 0.039 | 0.015 |
| 1000 | 0.041 | 0.083 | 0.036 | 0.036 | 0.071 | 0.036 |
| 2000 | 0.052 | 0.101 | 0.050 | 0.035 | 0.083 | 0.032 |

Global test: global hypothesis test with $\alpha = 5\%$.

Method 1: individual comparison with $\alpha = 5\%$.

Method 2: individual comparison with $\alpha = 5\%$ and Bonferroni.

Table 3. Powers obtained by applying the *EM-SEM* algorithms.

$PPV_1 = 0.90 \quad PPV_2 = 0.85 \quad NPV_1 = 0.80 \quad NPV_2 = 0.75$

$Se_1 = 0.943 \quad Se_2 = 0.944 \quad Sp_1 = 0.686 \quad Sp_2 = 0.50 \quad p = 75\%$

$\lambda_{11} = 0.50 \quad \lambda_{10} = 0.30 \quad \lambda_{01} = 0.30 \quad \lambda_{00} = 0.05$

| | $\alpha_1 = 1.03 \quad \alpha_0 = 1.50$ | | | $\alpha_1 = 1.04 \quad \alpha_0 = 1.75$ | | |
|---|---|---|---|---|---|---|
| n | Global test | Method 1 | Method 2 | Global test | Method 1 | Method 2 |
| 50   | 0.002 | 0.003 | 0.002 | 0.001 | 0.002 | 0.001 |
| 100  | 0.008 | 0.047 | 0.018 | 0.006 | 0.033 | 0.015 |
| 200  | 0.133 | 0.241 | 0.152 | 0.154 | 0.316 | 0.182 |
| 500  | 0.663 | 0.669 | 0.558 | 0.819 | 0.830 | 0.750 |
| 1000 | 0.975 | 0.942 | 0.895 | 0.998 | 0.993 | 0.982 |
| 2000 | 1     | 0.999 | 0.998 | 1     | 1     | 1     |

$\lambda_{11} = 0.95 \quad \lambda_{10} = 0.75 \quad \lambda_{01} = 0.75 \quad \lambda_{00} = 0.30$

| | $\alpha_1 = 1.03 \quad \alpha_0 = 1.50$ | | | $\alpha_1 = 1.04 \quad \alpha_0 = 1.75$ | | |
|---|---|---|---|---|---|---|
| n | Global test | Method 1 | Method 2 | Global test | Method 1 | Method 2 |
| 50   | 0.003 | 0.014 | 0.005 | 0.001 | 0.006 | 0.001 |
| 100  | 0.057 | 0.148 | 0.081 | 0.036 | 0.171 | 0.069 |
| 200  | 0.307 | 0.453 | 0.328 | 0.407 | 0.585 | 0.435 |
| 500  | 0.841 | 0.905 | 0.831 | 0.962 | 0.986 | 0.954 |
| 1000 | 0.998 | 1     | 0.998 | 0.999 | 0.999 | 0.999 |
| 2000 | 1     | 1     | 1     | 1     | 1     | 1     |

Global test: global hypothesis test with $\alpha = 5\%$.

Method 1: individual comparison with $\alpha = 5\%$.

Method 2: individual comparison with $\alpha = 5\%$ and Bonferroni.

Table 4. Type I errors obtained by applying multiple imputation.

$PPV_1 = PPV_2 = 0.85$  $NPV_1 = NPV_2 = 0.95$

$Se_1 = Se_2 = 0.85$  $Sp_1 = Sp_2 = 0.95$  $p = 25\%$

$\lambda_{11} = 0.50$  $\lambda_{10} = 0.30$  $\lambda_{01} = 0.30$  $\lambda_{00} = 0.05$

$\alpha_1 = 1.09$  $\alpha_0 = 10.50$

| | Global test | | | Individual tests with $\alpha = 5\%$ | | | Individual tests with Bonferroni | | |
|---|---|---|---|---|---|---|---|---|---|
| $n$ | Wald | Combined | LRT | L1 | W1 | K1 | L2 | W2 | K2 |
| 50 | 0 | 0 | 0 | 0.002 | 0 | 0 | 0.001 | 0 | 0 |
| 100 | 0 | 0 | 0 | 0.008 | 0 | 0.001 | 0.001 | 0 | 0 |
| 200 | 0.001 | 0.006 | 0.002 | 0.058 | 0.002 | 0.005 | 0.014 | 0.001 | 0.001 |
| 500 | 0.003 | 0.028 | 0.017 | 0.201 | 0.031 | 0.050 | 0.125 | 0.011 | 0.024 |
| 1000 | 0.009 | 0.031 | 0.032 | 0.268 | 0.054 | 0.069 | 0.183 | 0.010 | 0.032 |
| 2000 | 0.010 | 0.046 | 0.045 | 0.318 | 0.072 | 0.085 | 0.220 | 0.022 | 0.046 |

$\alpha_1 = 1.13$  $\alpha_0 = 15.25$

| | Global test | | | Individual tests with $\alpha = 5\%$ | | | Individual tests with Bonferroni | | |
|---|---|---|---|---|---|---|---|---|---|
| $n$ | Wald | Combined | LRT | L1 | W1 | K1 | L2 | W2 | K2 |
| 50 | 0 | 0 | 0 | 0 | 0 | 0 | 0 | 0 | 0 |
| 100 | 0 | 0 | 0 | 0 | 0 | 0 | 0 | 0 | 0 |
| 200 | 0 | 0.002 | 0 | 0.006 | 0.002 | 0.002 | 0.003 | 0.002 | 0.001 |
| 500 | 0.001 | 0.009 | 0.003 | 0.101 | 0.010 | 0.018 | 0.040 | 0.003 | 0.006 |
| 1000 | 0.002 | 0.019 | 0.009 | 0.180 | 0.033 | 0.040 | 0.108 | 0.002 | 0.014 |
| 2000 | 0.009 | 0.032 | 0.037 | 0.283 | 0.075 | 0.092 | 0.197 | 0.026 | 0.044 |

$\lambda_{11} = 0.95$  $\lambda_{10} = 0.75$  $\lambda_{01} = 0.75$  $\lambda_{00} = 0.30$

$\alpha_1 = 1.09$  $\alpha_0 = 10.50$

| | Global test | | | Individual tests with $\alpha = 5\%$ | | | Individual tests with Bonferroni | | |
|---|---|---|---|---|---|---|---|---|---|
| $n$ | Wald | Combined | LRT | L1 | W1 | K1 | L2 | W2 | K2 |
| 50 | 0 | 0 | 0 | 0.001 | 0 | 0 | 0 | 0 | 0 |
| 100 | 0 | 0 | 0 | 0.017 | 0.007 | 0.008 | 0.005 | 0 | 0.001 |
| 200 | 0.002 | 0.010 | 0.006 | 0.058 | 0.013 | 0.019 | 0.013 | 0.005 | 0.006 |
| 500 | 0.002 | 0.022 | 0.019 | 0.082 | 0.041 | 0.057 | 0.042 | 0.007 | 0.018 |
| 1000 | 0.003 | 0.045 | 0.042 | 0.123 | 0.065 | 0.082 | 0.065 | 0.020 | 0.034 |
| 2000 | 0.004 | 0.042 | 0.040 | 0.129 | 0.076 | 0.089 | 0.073 | 0.016 | 0.044 |

$\alpha_1 = 1.13$  $\alpha_0 = 15.25$

| | Global test | | | Individual tests with $\alpha = 5\%$ | | | Individual tests with Bonferroni | | |
|---|---|---|---|---|---|---|---|---|---|
| $n$ | Wald | Combined | LRT | L1 | W1 | K1 | L2 | W2 | K2 |
| 50 | 0 | 0 | 0 | 0 | 0 | 0 | 0 | 0 | 0 |
| 100 | 0 | 0 | 0 | 0.001 | 0 | 0 | 0 | 0 | 0 |
| 200 | 0 | 0.001 | 0.001 | 0.019 | 0.009 | 0.013 | 0.004 | 0 | 0 |
| 500 | 0 | 0.014 | 0.011 | 0.069 | 0.031 | 0.037 | 0.029 | 0.007 | 0.012 |
| 1000 | 0 | 0.020 | 0.024 | 0.094 | 0.047 | 0.059 | 0.045 | 0.020 | 0.024 |
| 2000 | 0.002 | 0.049 | 0.044 | 0.126 | 0.074 | 0.085 | 0.067 | 0.024 | 0.047 |

Wald: Global Wald test. Combined: Combination of p-values. LRT: Combined likelihood-ratio tests.

Method $L1$: individual comparison applying the method of Leisenring et al with $\alpha = 5\%$.

Method $W1$: individual comparison applying the method of Wang et al with $\alpha = 5\%$.

Method $K1$: individual comparison applying the method of Kosinski with $\alpha = 5\%$.

Method $L2$: individual comparison applying the method of Leisenring et al with $\alpha = 5\%$ and Bonferroni.

Method $W2$: individual comparison applying the method of Wang et al with $\alpha = 5\%$ and Bonferroni.

Method $K2$: individual comparison applying the method of Kosinski with $\alpha = 5\%$ and Bonferroni.

Table 5. Powers obtained by applying multiple imputation.

$PPV_1 = 0.90 \quad PPV_2 = 0.85 \quad NPV_1 = 0.80 \quad NPV_2 = 0.75$

$Se_1 = 0.943 \quad Se_2 = 0.944 \quad Sp_1 = 0.686 \quad Sp_2 = 0.50 \quad p = 75\%$

$\lambda_{11} = 0.50 \quad \lambda_{10} = 0.30 \quad \lambda_{01} = 0.30 \quad \lambda_{00} = 0.05$

$\alpha_1 = 1.03 \quad \alpha_0 = 1.50$

| | Global test | | | Individual tests with $\alpha = 5\%$ | | | Individual tests with Bonferroni | | |
|---|---|---|---|---|---|---|---|---|---|
| $n$ | Wald | Combined | LRT | L1 | W1 | K1 | L2 | W2 | K2 |
| 50 | 0 | 0.001 | 0 | 0.009 | 0.002 | 0.002 | 0.001 | 0 | 0 |
| 100 | 0.002 | 0.024 | 0.017 | 0.128 | 0.040 | 0.039 | 0.069 | 0.013 | 0.013 |
| 200 | 0.005 | 0.248 | 0.135 | 0.468 | 0.235 | 0.229 | 0.367 | 0.154 | 0.149 |
| 500 | 0.065 | 0.762 | 0.573 | 0.887 | 0.690 | 0.691 | 0.834 | 0.577 | 0.576 |
| 1000 | 0.341 | 0.980 | 0.908 | 0.995 | 0.934 | 0.935 | 0.976 | 0.887 | 0.886 |
| 2000 | 0.723 | 1 | 1 | 1 | 0.999 | 0.999 | 1 | 0.998 | 0.998 |

$\alpha_1 = 1.04 \quad \alpha_0 = 1.75$

| | Global test | | | Individual tests with $\alpha = 5\%$ | | | Individual tests with Bonferroni | | |
|---|---|---|---|---|---|---|---|---|---|
| $n$ | Wald | Combined | LRT | L1 | W1 | K1 | L2 | W2 | K2 |
| 50 | 0 | 0 | 0 | 0.002 | 0.001 | 0.001 | 0.001 | 0 | 0 |
| 100 | 0.003 | 0.014 | 0.005 | 0.119 | 0.029 | 0.024 | 0.049 | 0.010 | 0.010 |
| 200 | 0.008 | 0.334 | 0.175 | 0.520 | 0.279 | 0.272 | 0.393 | 0.170 | 0.164 |
| 500 | 0.082 | 0.912 | 0.761 | 0.940 | 0.806 | 0.803 | 0.910 | 0.722 | 0.713 |
| 1000 | 0.349 | 0.998 | 0.986 | 1 | 0.989 | 0.989 | 1 | 0.972 | 0.971 |
| 2000 | 0.725 | 1 | 1 | 1 | 0.999 | 0.999 | 1 | 0.998 | 0.998 |

$\lambda_{11} = 0.95 \quad \lambda_{10} = 0.75 \quad \lambda_{01} = 0.75 \quad \lambda_{00} = 0.30$

$\alpha_1 = 1.03 \quad \alpha_0 = 1.50$

| | Global test | | | Individual tests with $\alpha = 5\%$ | | | Individual tests with Bonferroni | | |
|---|---|---|---|---|---|---|---|---|---|
| $n$ | Wald | Combined | LRT | L1 | W1 | K1 | L2 | W2 | K2 |
| 50 | 0.003 | 0.004 | 0.004 | 0.039 | 0.024 | 0.020 | 0.015 | 0.007 | 0.002 |
| 100 | 0.031 | 0.088 | 0.075 | 0.244 | 0.178 | 0.170 | 0.149 | 0.102 | 0.080 |
| 200 | 0.118 | 0.343 | 0.318 | 0.539 | 0.482 | 0.473 | 0.431 | 0.346 | 0.331 |
| 500 | 0.409 | 0.861 | 0.833 | 0.915 | 0.891 | 0.889 | 0.872 | 0.834 | 0.831 |
| 1000 | 0.747 | 1 | 1 | 1 | 1 | 1 | 1 | 1 | 1 |
| 2000 | 1 | 1 | 1 | 1 | 1 | 1 | 1 | 1 | 1 |

$\alpha_1 = 1.04 \quad \alpha_0 = 1.75$

| | Global test | | | Individual tests with $\alpha = 5\%$ | | | Individual tests with Bonferroni | | |
|---|---|---|---|---|---|---|---|---|---|
| $n$ | Wald | Combined | LRT | L1 | W1 | K1 | L2 | W2 | K2 |
| 50 | 0.001 | 0.001 | 0.001 | 0.018 | 0.012 | 0.006 | 0.003 | 0.001 | 0.001 |
| 100 | 0.011 | 0.071 | 0.056 | 0.232 | 0.183 | 0.170 | 0.134 | 0.095 | 0.077 |
| 200 | 0.074 | 0.458 | 0.412 | 0.665 | 0.588 | 0.578 | 0.539 | 0.454 | 0.436 |
| 500 | 0.384 | 0.961 | 0.948 | 0.983 | 0.975 | 0.974 | 0.966 | 0.951 | 0.948 |
| 1000 | 0.669 | 1 | 1 | 1 | 1 | 1 | 1 | 1 | 1 |
| 2000 | 1 | 1 | 1 | 1 | 1 | 1 | 1 | 1 | 1 |

Wald: Global Wald test. Combined: Combination of p-values. LRT: Combined likelihood-ratio tests.

Method $L1$: individual comparison applying the method of Leisenring et al with $\alpha = 5\%$.

Method $W1$: individual comparison applying the method of Wang et al with $\alpha = 5\%$.

Method $K1$: individual comparison applying the method of Kosinski with $\alpha = 5\%$.

Method $L2$: individual comparison applying the method of Leisenring et al with $\alpha = 5\%$ and Bonferroni.

Method $W2$: individual comparison applying the method of Wang et al with $\alpha = 5\%$ and Bonferroni.

Method $K2$: individual comparison applying the method of Kosinski with $\alpha = 5\%$ and Bonferroni.

Table 6. Relative biases of estimators of predictive values.

| | $PPV_1 = PPV_2 = 0.80$ $NPV_1 = NPV_2 = 0.90$ | | | | | | |
|---|---|---|---|---|---|---|---|
| | $Se_1 = Se_2 = 0.914$ $Sp_1 = Sp_2 = 0.771$ $p = 50\%$ | | | | | | |
| | $\lambda_{11} = 0.50$ $\lambda_{10} = 0.30$ $\lambda_{01} = 0.30$ $\lambda_{00} = 0.05$ | | | | | | |
| | $\alpha_1 = 1.05$ $\alpha_0 = 2.68$ | | | | $\alpha_1 = 1.07$ $\alpha_0 = 3.53$ | | |
| | EM-SEM | | MICE | | EM-SEM | | MICE | |
| $n$ | $\hat{PPV}_1$ | $\hat{NPV}_1$ | $\overline{PPV}_1$ | $\overline{NPV}_1$ | $\hat{PPV}_1$ | $\hat{NPV}_1$ | $\overline{PPV}_1$ | $\overline{NPV}_1$ |
| 50   | -0.144 | -0.293 | -0.160 | -0.264 | -0.141 | -0.301 | -0.158 | -0.252 |
| 100  | -0.074 | -0.211 | -0.085 | -0.192 | -0.073 | -0.217 | -0.084 | -0.179 |
| 200  | -0.041 | -0.134 | -0.047 | -0.128 | -0.037 | -0.133 | -0.042 | -0.123 |
| 500  | -0.018 | -0.065 | -0.021 | -0.077 | -0.015 | -0.054 | -0.018 | -0.066 |
| 1000 | -0.008 | -0.031 | -0.010 | -0.045 | -0.008 | -0.027 | -0.011 | -0.045 |
| 2000 | -0.004 | -0.020 | -0.007 | -0.026 | -0.004 | -0.019 | -0.006 | -0.026 |
| | $\lambda_{11} = 0.95$ $\lambda_{10} = 0.75$ $\lambda_{01} = 0.75$ $\lambda_{00} = 0.30$ | | | | | | |
| | $\alpha_1 = 1.05$ $\alpha_0 = 2.68$ | | | | $\alpha_1 = 1.07$ $\alpha_0 = 3.53$ | | |
| | EM-SEM | | MICE | | EM-SEM | | MICE | |
| $n$ | $\hat{PPV}_1$ | $\hat{NPV}_1$ | $\overline{PPV}_1$ | $\overline{NPV}_1$ | $\hat{PPV}_1$ | $\hat{NPV}_1$ | $\overline{PPV}_1$ | $\overline{NPV}_1$ |
| 50   | -0.101 | -0.143  | -0.103 | -0.150 | -0.101 | -0.136 | -0.101 | -0.141 |
| 100  | -0.047 | -0.0750 | -0.048 | -0.083 | -0.050 | -0.073 | -0.051 | -0.080 |
| 200  | -0.023 | -0.039  | -0.025 | -0.047 | -0.024 | -0.032 | -0.025 | -0.038 |
| 500  | -0.009 | -0.016  | -0.010 | -0.021 | -0.010 | -0.015 | -0.011 | -0.019 |
| 1000 | -0.005 | -0.008  | -0.007 | -0.013 | -0.004 | -0.008 | -0.005 | -0.011 |
| 2000 | -0.003 | -0.004  | -0.004 | -0.007 | -0.003 | -0.006 | -0.004 | -0.008 |

Table 7. Data from the study of Hall et al and results.

| | Observed frequencies | | | | |
|---|---|---|---|---|---|
| | $T_1 = 1$ | | $T_1 = 0$ | | |
| | $T_2 = 1$ | $T_2 = 0$ | $T_2 = 1$ | $T_2 = 0$ | Total |
| $V = 1$ | | | | | |
| $D = 1$ | 31 | 5 | 3 | 1 | 40 |
| $D = 0$ | 25 | 10 | 19 | 55 | 109 |
| $V = 0$ | 22 | 6 | 65 | 346 | 439 |
| Total | 78 | 21 | 87 | 402 | 588 |
| | Results | | | | |
| | $\hat{P}PV_1 \pm SE$ | $\hat{N}PV_1 \pm SE$ | $\hat{P}PV_2 \pm SE$ | $\hat{N}PV_2 \pm SE$ | |
| EM-SEM | $0.507 \pm 0.059$ | $0.961 \pm 0.020$ | $0.334 \pm 0.052$ | $0.966 \pm 0.018$ | |
| MICE | $0.504 \pm 0.062$ | $0.948 \pm 0.021$ | $0.327 \pm 0.052$ | $0.949 \pm 0.020$ | |